\documentclass[aps,prb,showpacs,floatfix,amsmath,amssymb,superscriptaddress,reprint]{revtex4-1}
\usepackage{float}
\usepackage{tikz}
\usepackage{bbold}
\usepackage[normalem]{ulem}
\usepackage{color}
\usepackage{graphicx}
\usepackage{natbib}
\usepackage{epsfig}
\usepackage{setspace}
\usepackage{amsmath}
\usepackage{amssymb}
\usepackage{verbatim}
\usepackage{bibentry}
\usepackage{hyperref}
\usepackage{cancel}
\usepackage{xcolor}

\definecolor{scarred}{rgb}{0.75,0.0,0.0}
\graphicspath{ {./Images/} }
\usepackage{lineno}
\usepackage{enumitem} 

\begin{document}
\title{Emergent soft-gap Anderson models at quantum criticality in a lattice Hamiltonian within dynamical mean field theory}
\author{Sujan K. K.}
\affiliation{Jawaharlal Nehru Centre for Advanced Scientific Research, Jakkur PO, Bengaluru 560064, India.}
\author{Vinayak M.\ Kulkarni}
\affiliation{Jawaharlal Nehru Centre for Advanced Scientific Research, Jakkur PO, Bengaluru 560064, India.}
\author{N.\ S.\ Vidhyadhiraja}\email{raja@jncasr.ac.in}
\affiliation{Jawaharlal Nehru Centre for Advanced Scientific Research, Jakkur PO, Bengaluru 560064, India.}
\author{Sudeshna Sen}\email{sudeshna@iitism.ac.in}
\affiliation{Department of Physics, Indian Institute of Technology, Dhanbad.}
\begin{abstract}
\noindent
   Local quantum criticality in itinerant fermion systems has been extensively investigated through the soft-gap Anderson impurity model, wherein a localized, correlated impurity, hybridizes with a broad conduction band with a singular, $|\omega|^r$, density of states. 
   However, lattice models hosting quantum critical points (QCPs), do not appear to have such a spectrum emerging at the QCP. 
   In this work, we report the emergence of such a singular form of the density of states in a three-orbital lattice model, within dynamical mean field theory, precisely at a quantum critical point, separating a gapless, Fermi liquid, metallic phase from a gapped, Mott insulating phase. 
   A temperature-dependent exponent, $\alpha$, defined using the corresponding Matsubara self-energy, is found to vary from $+1$ deep in the FL regime, to $-1$ in the Mott insulator regime. 
   Interestingly, we find that $\alpha$ becomes temperature independent, and hence isosbestic, precisely at the QCP. 
   The isosbestic exponent is shown to lead to an emergent soft-gap spectrum, $|\omega|^r$ at the QCP, where $r = |\alpha_{\rm iso}|$. 
   We discuss the implications of our findings for non-Fermi liquid behaviour in the quantum critical region of the phase diagram. 
\end{abstract}
\maketitle
\noindent
\section{Introduction}
\label{sec:intro}
Quantum phases and their critical nature continues to be a leading field of study in condensed matter physics~\cite{coleman2005quantum}. 
Particularly challenging and interesting is the class of quantum phase transitions (QPTs) that defy the traditional Landau-Ginzburg-Wilson framework~\cite{sachdev2010quantum,Vojta_imp_QPT,Vojta_2003,Si_Steglich_2010}. 
These include topological phase transitions~\cite{top_on_top,continentino2017topological,brink2018topological,maric2020quantum} and QPTs involving the presence of both locally critical modes and long wavelength fluctuations~\cite{Vojta_imp_QPT,Si_Steglich_2010}, commonly observed in heavy fermion systems~\cite{gegenwart2008quantum,si2006global,neumann2007bilayer}. 
Other systems hosting local quantum critical points (QCPs) are correlated impurity systems with a pseudogapped host~\cite{Vojta_imp_QPT,fritz2013physics,mitchell2013kondo,Anders-2-channel-SGAM} and lattice Dirac systems in two dimensions~\cite{li2017fermion,Meng1,Mott_MI_in_Dirac_sys_Meng2,FRG_Dirac_system,boyack2021quantum,Scalettar_Dirac_fermion, Dirac_Assad, Weyl_QPT}. 
For example, in the soft-gap Anderson model (SGAM), the local (boundary) QCP is governed by the intrinsic critical local moment fluctuations that are further sensitized in the presence of the vanishingly small host density of states~\cite{gonzalez1996stabilization,chen1995kondo,ingersent1996behavior,bulla1997anderson,bulla2000soft,Vojta_imp_QPT,logan2000local}. 
From a field theoretical perspective, local QCPs represent interacting fixed points when the exponent, $r$, of the vanishing DoS is less than one~\cite{Vojta_imp_QPT,ingersent2002critical}. 

An intriguing scenario would arise if such gapless fermionic excitations naturally emerge precisely at the QCP instead of being existent from the outset. 
For example, the deconfined Mott transition is associated with the emergence of a gapless spin liquid state with a spinon Fermi surface representative of neutral spin-$1/2$ quasi-particles while the charge excitations are gapped\cite{senthil2004deconfined}. 
Disorder induced power-law gaps have also been observed across an insulator-metal Mott transition~\cite{wang2018disorder}.
Identifying new models and the underlying universality classes will broaden our current understanding of QPTs beyond the Landau-Ginzburg-Wilson framework. 
Particularly enlightening would be the expansion of this class of local QPTs representing metal-insulator transitions, across naturally occurring lattice settings to aid experimental verification. 
In principle, the paramagnetic metal to paramagnetic insulator Mott transition could represent such fermionic criticality via the emergence of scale invariant power law spectra~\cite{Imada_MIT_QCP,Vojta_pseudogap_HM,imada1998metal}. 
However, as shown in a recent study a strictly power law spectrum down to zero temperature has not been observed in the plain vanilla Hubbard model~\cite{Vojta_pseudogap_HM}. 
Concomitantly, the critical nature of the Mott transition is indeed observable, but only beyond a typical finite temperature scale~\cite{vuvcivcevic2013finite}. 
A recent experiment on M\"oire lattice also demonstrated a continuous Mott transition ~\cite{li2021continuous}.
Theoretically, a strictly zero temperature, continuous Mott transition was indeed observed in a three orbital lattice model consisting of a localised, flat band of interacting ($f$) electrons hybridising with two independent bands of itinerant electrons ($c$)\cite{sen2016quantum}. 
This model has recently been studied in other contexts and can also be viewed as a modified periodic Anderson model in a layered configuration~\cite{hu2017effects,dos2020superconducting,yang2019degenerate}. 

In this work, we revisit this model to analyse the signatures of quantum criticality at non-zero temperatures. 
We demonstrate the pristine nature of a continuous Mott transition in this model within dynamical mean field theory (DMFT)~\cite{dmft_rmp} using numerically exact continuous time quantum Monte Carlo method~\cite{seth2016triqs, werner2006continuous,werner2006hybridization,gull2008continuous,boehnke2011orthogonal,boehnke2015susceptibilities,CTQMC_review}. 
The QPT occurs as the quasiparticle excitations continuously evolve into non-Fermi liquid and critical gapless states manifested in the form of a power law spectrum down to the lowest energy scales. 
Specifically, the critical spectra represent emergent soft-gap Anderson models with a characteristic exponent for the soft-gap density of states. 
Interestingly, this exponent is independent of temperature over a given QCP, but is non-universal, and varies over the line of quantum critical points.

The paper is organised as follows: the model and method is presented in Section~\ref{sec:model} followed by Results and Discussions in Section~\ref{sec:results}. Here we identify the quantum critical points and demonstrate the presence of an isosbestic exponent and quantum criticality through the analysis of the low frequency form of the self-energy. We also relate these observations to an emergent soft-gap density of states and hybridization function at the QCP. In Section~\ref{sec:outlook}, we conclude and discuss some future directions.

\noindent
\section{Model and formalism}
\label{sec:model}
The three-orbital model system\cite{sen2016quantum} comprises a localised, correlated $f$ orbital hybridising with a conduction band ($c$) of itinerant electrons via the hybridization energy, $V$; the latter couples to a third, delocalised, non-interacting ($M$) orbital via hopping energy, $t_\perp$. The Hamiltonian, $H$, is thus given by, $H = H_f + H_c + H_M + H_{fc} + H_{cM}$ and the individual terms are represented in standard second quantized notation as :
\begin{align}
    H_f & = \epsilon_f\sum_{i\sigma} f^\dag_{i\sigma}f^{\phantom{\dag}}_{i\sigma} + U\sum_i n_{fi\uparrow}n_{fi\downarrow} \\
    H_c + H_M & =\sum_{k\sigma}\epsilon_k (c_{k\sigma}^\dagger
c_{k\sigma}^{\phantom{\dagger}} + c_{Mk\sigma}^\dagger
c_{Mk\sigma}^{\phantom{\dagger}})\\
H_{fc} & = V \sum_{k\sigma}( f^\dag_{k\sigma}c_{k\sigma}^{\phantom{\dagger}} + {\rm h.c})\\
H_{cM} &= t_\perp \sum_{k\sigma}( c^\dag_{k\sigma}c_{Mk\sigma}^{\phantom{\dagger}} + {\rm h.c}  )\,.
\end{align}
In the absence of $c$-$M$ mixing ($t_\perp=0$), this is simply a standard periodic Anderson model, that has been extensively employed to investigate heavy fermion systems~\cite{dmft_rmp,peters2006numerical,jarrell1993periodic,vidhyadhiraja2004dynamics,vidhyadhiraja2005optical,logan2005dynamics}.
In a previous study \cite{sen2016quantum}, two of the authors had employed the local moment approach\cite{logan1998local} within DMFT at zero temperature, and had identified a surface of quantum critical points, in the $U-t_\perp-V$ space, separating a gapless Fermi liquid phase from a Mott insulator phase. 
For a fixed-$V$, the line of critical points was found to be roughly hyperbolic in the $t_\perp-U$ plane, with the asymptotes being $t_\perp=1$ and $U=0$. 
However, the finite temperature signatures of the critical transition and underlying mechanism of the QCP were not addressed.
Hence, in this work, we use continuous time quantum Monte-Carlo, as implemented in the TRIQS open source package~\cite{seth2016triqs, werner2006continuous,werner2006hybridization,gull2008continuous,boehnke2011orthogonal,boehnke2015susceptibilities,CTQMC_review} to investigate the nature of this quantum criticality and the manifestation of the zero temperature QCPs in the finite temperature phase diagram. 
We work in the particle-hole symmetric, paramagnetic regime, with $\epsilon_f=-U/2$, thus avoiding the fermion sign problem. 

The local $f$-Green's function, in this model, is given by, 
\begin{equation}
    G_{f}(i\omega_n)= \int\,d\epsilon\,\rho_0(\epsilon) \frac{1}{i\omega_n
- \frac{V^2}{i\omega_n - \epsilon - \frac{t_\perp^2}{i\omega_n-\epsilon}}-\Sigma(i\omega_n)}\,,\label{eq:gfeq}
\end{equation} where $\Sigma(i\omega_n)$ is the $k$-independent (within DMFT) self-energy~\cite{sen2016quantum}.
We choose a generic, semi-elliptic, conduction band density of states, $\displaystyle \rho_0(\epsilon) = 2\sqrt{1 - \epsilon^2/D^2}/\pi D$ with $D$ being the half band-width. 
It is important to note the non-singular choice of the density of states, which is in contrast to the choice made in the soft-gap Anderson model ($|\epsilon|^r$). 
The Weiss field, $\Delta(i\omega_n)$,  is the self-consistent hybridization function representing the DMFT bath for the $f$-electrons, and is related to the host Green's function, $\mathcal{G}(i\omega_n)$, by, $\Delta(i\omega_n) = i\omega_n -\mathcal{G}^{-1}(i\omega_n)$, which in turn can be obtained from the  self energy, $\Sigma_f(i\omega_n)$, and the $f$-Greens function through the Dyson's equation $\mathcal{G}^{-1}(i\omega_n) = G_{f}^{-1}(i\omega_n)+\Sigma_f(i\omega_n)$. 
Thus, for a given $\Delta(i\omega_n)$, we compute the self-energy and Green's function of the impurity using a hybridization-expansion CTQMC as implemented in TRIQS~\cite{seth2016triqs, werner2006continuous,werner2006hybridization,gull2008continuous,boehnke2011orthogonal,boehnke2015susceptibilities,CTQMC_review}. 
We use the Pad\'{e} approximant method\cite{schott2016analytic} to analytically continue the Matsubara $f$-Green's function to real frequencies, and hence obtain the real-frequency $f$-spectrum.
In this work, we use a fixed $V=0.44D$ and two representative interaction strengths, namely, $U=1.75D,\,1.2D$, with the unit of energy being $D=1$ to analyse the phase transition as a function of $t_\perp$. In fact, it should be noted that for each $V$, this model exhibits a line of quantum critical points in the $t_\perp\,-\,U$ plane~\cite{sen2016quantum}. The parameter sets used in this work to study the finite temperature signatures of the QPT driven by tuning $t_\perp$ thus represent only two such points on a surface of QCPs existing in this model. 

\section{Results and Discussion}
\label{sec:results}
\subsection{Identification of quantum critical points}
\label{subsec:identify_QCP}
\begin{figure}[tbh]
    \centering
    
    \includegraphics[clip=,width=1.0\columnwidth]{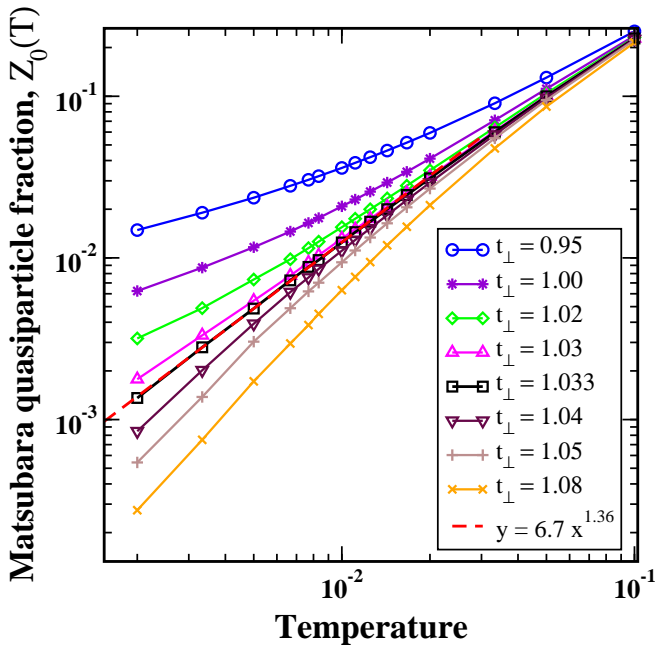}
    \caption{The Matsubara quasiparticle fraction, $Z_0(T)$ defined as, $Z_0(T)=\left[1-\mathrm{Im} \Sigma(i\omega_0)/\omega_0\right]^{-1}$, calculated using CTQMC, is plotted as a function of temperature, $T$, for a selection of $t_\perp$ values. 
    The dashed red line is a power law fit to the $t_\perp=1.033$ data. The fit is seen to be valid over almost two decades.}
    \label{fig:qpwvsT}
\end{figure}

We utilise the temperature dependent, Matsubara quasiparticle fraction, $Z_0(T)=\left[1-\mathrm{Im} \Sigma(i\omega_0)/\omega_0\right]^{-1}$, with 
$\omega_0=\pi T$ being the lowest Matsubara frequency, for locating the QCP~\cite{vidhyadhiraja2009quantum,chen2012lifshitz}. 
In the zero temperature limit, $\lim_{T\to 0} Z_0(T)= Z$, is simply the zero temperature quasiparticle renormalization factor, defined for a conventional, real frequency, Fermi-liquid self energy, as $\Sigma_{FL} = \Sigma(0) + \omega(1-1/Z) + {\mathcal{O}}(\omega^2)$. Naturally, at finite temperatures, and over the Mott insulator regime, the $Z_0(T)$ does not have a direct physical interpretation. Nevertheless, it has proven to be a quick diagnostic for identifying a QCP~\cite{vidhyadhiraja2009quantum,pourovskii2013electronic,chen2012lifshitz,keen2020quantum}.
Considering $U$ and $V$ to be $1.75$ and $0.44$ (in units of $D$), respectively, and for various values of $t_\perp$, the computed $Z_0(T)$ are shown in Fig.~\ref{fig:qpwvsT}. We have performed calculations for many more $t_\perp$ values, but have not displayed the results to avoid cluttering.
The low $T$ Matsubara quasiparticle fraction, $Z_0(T)$ clearly changes its character as $t_\perp$ increases through $t_{\perp}/D=1.033$, in a way that parallels the finding in a two-dimensional Hubbard model close to optimal doping\cite{vidhyadhiraja2009quantum} that had employed dynamical cluster approximation framework\cite{maier2005quantum}.
The data for $t_\perp<t_{\perp c}$ has a convex curvature as $T\to 0$ and has a finite positive intercept at $t_{\perp c}$ while $Z_0(T)$ for $t_\perp>t_{\perp c}$ bends downwards displaying a concave curvature approaching a value close to zero at a finite temperature. 
We identify the QCP as the value of $t_\perp$ at which such a vivid change in the dependence of $Z_0$ on $T$ occurs; 
specifically at the parameter,  $t_{\perp c}/D=1.033\pm 0.001$ the $Z_0(T)\sim T^{1.36\pm0.005 }$ has a power law form. 
The behaviour indicates a crossover from a Fermi liquid phase for $t_\perp < t_{\perp c}$ to a Mott insulator phase for $t_\perp>t_{\perp c}$~\cite{sen2016quantum}.
A parallel analysis for $U=1.2D$ yields $t_{\perp,c}/D=1.077\pm 0.001 $ and $Z_0(T)\sim T^{1.38\pm 0.004}$~\cite{errorbar}.

In order to consolidate the finding of a quantum critical point, we must show that the characteristic scales on the two sides of the QCP vanish simultaneously implying the absence of a co-existence region. 
The scale on the Fermi liquid side can be taken to be the zero-temperature quasi-particle weight $Z_0(T=0)$. 
We extract this quantity as a $y$-axis intercept of a simple fit of the $Z_0(T)$ data (using a form $Z_0(0) + CT^\gamma$), for a range of $t_\perp < 1.033D$.  
On the Mott insulator side, a natural choice of the characteristic scale is the spectral gap, the determination of which requires analytical continuation, a mathematically ill-defined problem. 
As an alternative to analytic continuation from imaginary to real frequencies for determining the Mott spectral gap, we 
 use a protocol that employs the reverse route, which is well-defined, albeit approximate. This protocol is described briefly here, and details have been provided in Appendix~\ref{app1}. 
 
 A real frequency gapped density of states is {\em designed} with certain unknown parameters. Using three constraints, namely, the spectral sum rule, getting the density of states value at $\omega=0$ directly from CTQMC as $A(\omega=0)=-G_f(\tau=\beta/2)/(\pi T)$, and a geometric constraint between the parameters, we reduce the number of unknown parameters to just two, from which we can obtain the zero temperature gap, $\Delta_{g0}$, and an effective bandwidth, $W$.  We also obtain a finite temperature spectral gap, that we call $\Delta_{gT}$, such that at $T=0$, $\Delta_{gT}=\Delta_{g0}$. The spectral representation may then be used to transform the above two-parameter real-frequency density of states, to a Matsubara frequency Greens function, $G_f^{\rm model}$, having two fitting parameters, which can be found simply by comparing against the numerically obtained Greens function.
In Fig.~\ref{fig:app_fig2} of Appendix~\ref{app1} we demonstrate a sample of the fitting of the CTQMC data with the two-parameter model Greens function.
Utilizing the protocol described in Appendix~\ref{app1}, and data for the lowest temperature reached in our simulations, i.e, $T/D=0.002$,  we have computed the spectral gap scales ($\Delta_{gT}$ and $\Delta_{g0}$).

\begin{figure}[tbh]
\centering
   
    \includegraphics[clip=,scale=0.4]{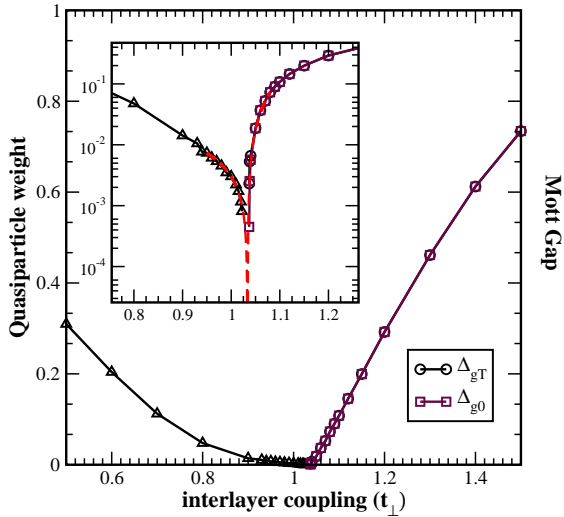}
    \caption{Vanishing of the zero temperature quasi-particle weight($Z$) and zero (finite) temperature Mott gap, denoted as $\Delta_{g0}$ ($\Delta_{gT}$) as we approach the QCP ($t_{\perp,c} \approx 1.033$) plotted on a linear scale (main panel) and evaluated at $U/D=1.75$ and $T/D=0.002$. (Inset) The same data in the main panel is shown on a linear-log scale for the region close to the QCP to highlight that the scales vanish as a power law, with $Z\sim |t-t_{\perp c}|^{a_1}$ and $\Delta_{g0}\sim |t-t_{\perp c}|^{a_2}$ (with $a_1=0.97\pm0.02$, $a_2=1.35\pm0.02$). The power-law fit to the CTQMC data is shown as red dashed-lines. The Fermi liquid scale $Z$ and the zero-temperature Mott gap $\Delta_{g0}$ vanish simultaneously, at $t_{\perp c}\approx1.033D$, which thus represents a QCP.}
    \label{fig:energyscales}
\end{figure}

In Fig.~\ref{fig:energyscales} we show the quasiparticle weight, $Z\equiv Z_0 (T=0)$, and the MI gap scale $\Delta_{g0}$ (as well as $\Delta_{gT}$) for $U/D=1.75$, as a function of $t_\perp$ on a linear scale in the main panel and on a linear-log scale in the inset.  The zero and finite temperature Mott gaps ($\Delta_{g0}$ and $\Delta_{gT}$ respectively) are almost identical for all $t_\perp$ except in the proximity of the QCP, where they do differ. 
We have confirmed that the two scales, namely $Z$ and $\Delta_{g0}$, vanish simultaneously as a power law with $Z\sim |t_\perp-t_{\perp c}|^{a_1}$ and $\Delta_{g0}\sim |t_\perp-t_{\perp c}|^{a_2}$ ($a_1=0.97\pm0.02,\, a_2=1.35\pm0.02$)~\cite{errorbar} at $t_{\perp c}\approx1.033D$ thus representing a QCP at $t_{\perp c}$. Again, for the other parameter set, namely $U=1.2D, t_{\perp,c}=1.077D$, we find a similar vanishing of scales with $a_1,a_2=0.89\pm0.02,1.01\pm0.01$~\cite{errorbar}, confirming that this parameter set is also a QCP.

\noindent
\subsection{Isosbestic points and quantum criticality}
\label{subsec:isosbestic_points}
\begin{figure}[tbh]
\centering
   
    \includegraphics[clip=,width=\columnwidth]{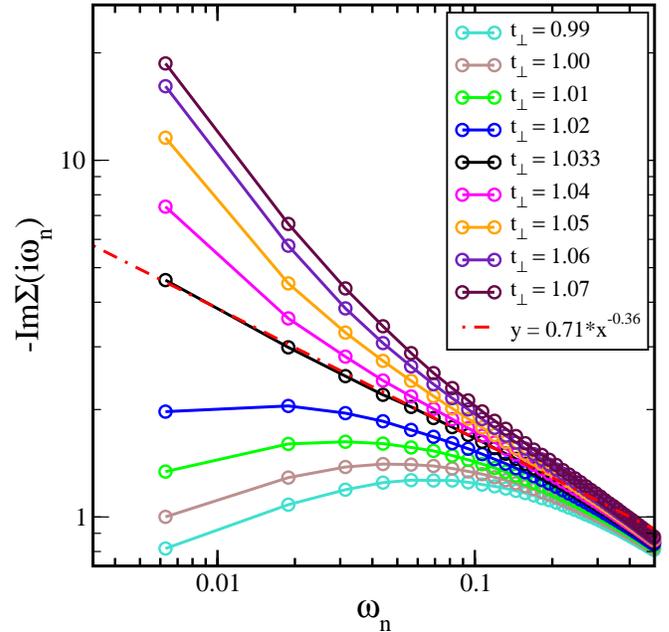}
    \caption{The imaginary part of the Matsubara freqency self-energy, $\Sigma(i\omega_n)$, as a function of Matsubara frequency, $\omega_n$, for various $t_{\perp}$ values (in units of $D$) evaluated at $T/D=0.002$, $U/D=1.75$, and $V/D=0.44$. The power law ($\sim |\omega_n|^{-|\alpha|}$) form is observed at the QCP ($t_{\perp,c} \approx 1.033D$) with an exponent $\alpha = - 0.36\pm0.005$.~\cite{errorbar}} 
    \label{fig:imsigma}
\end{figure}

For an ideal Fermi liquid and a Mott insulator,  the imaginary part of the Matsubara self-energy may be represented as $-{\rm Im}\Sigma(i\omega_n)\sim (\omega_n)^{\alpha}$ for $\omega_n\to 0^+$, with $\alpha=+1$ and $-1$ respectively \cite{logan2015mott}. 
However, if the respective metal-insulator transition is a quantum-critical phase transition, then quantum-critical fluctuations at finite temperatures generally accompany classic non-Fermi liquid signatures in the form of low-frequency power law behaviour of the self-energy. As we move farther away from the QCP, these non-Fermi liquid features crossover to the conventional features representing the stable metallic or insulating ground states on either side of the QCP. In order to understand whether our system projects similar characteristics, we plot $-{\rm Im}\Sigma(i\omega_n)$ as a function of $\omega_n$ on a log-log scale for various values of $t_\perp$, evaluated at $U/D=1.75$ and $T/D=0.002$. As shown in Fig.\ref{fig:imsigma}, and as expected, a power-law behaviour extending over a wide range of frequencies is observed at $t_{\perp c}\approx 1.033D$. 

As we move away from this parameter the $-{\rm Im}\Sigma(i\omega_n)$ deviates from its power law characteristic, especially at low frequencies.
The magnitude of this deviation signifies the parametric distance from the QCP. 
We define a self-energy exponent, $\alpha$, using a two-parameter fit of $y_n=-{\rm Im}\Sigma(i\omega_n,t_\perp,T)$ at a fixed $t_\perp$ and $T$ fitting only the first two Matsubara frequencies to $A|\omega_n|^\alpha$.  The exponent may be obtained in closed form as $\alpha=\ln(y_1/y_0)/\ln(3)$. The Matsubara quasiparticle fraction, $Z_0(T)$ acquires a physical meaning as the quasiparticle weight only at $T=0$. Similarly, the exponent $\alpha$ acquires a physical meaning as the exponent of the emergent soft-gap DoS only at the QCP (as will be shown later). 
Nevertheless, as the change of curvature in the temperature dependence of $Z_0(T)$ leads to an identification of a QCP, the exponent $\alpha$, as shown in the following, turns out to be an excellent diagnostic for identifying the QCP.

\begin{figure}[hbt]
   
   \includegraphics[clip=,width=1.0\columnwidth]{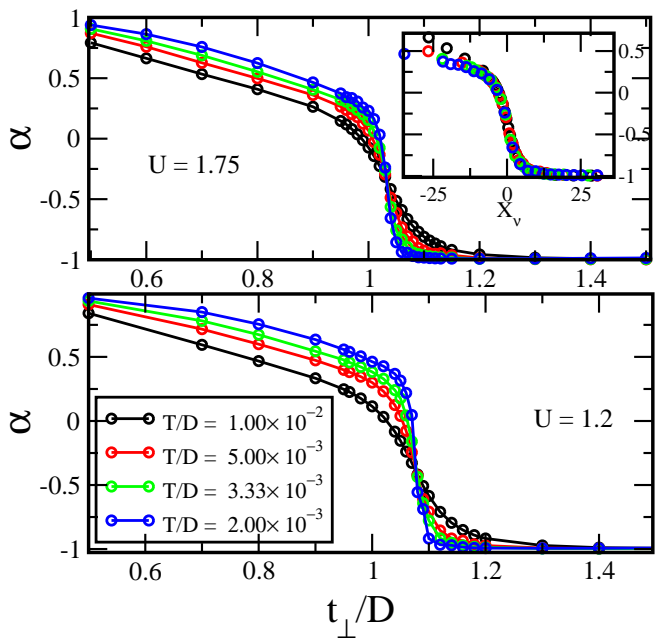}
\caption{The self-energy exponent $\alpha$ (defined in the text through the imaginary part of the self-energy) is plotted as a function of $t_{\perp}/D$ for (a) $U/D = 1.75$ (left panel) and (b) and $U/D = 1.2$ (right panel) for different $T/D$. 
The exponent features an isosbestic point, at $t_{\perp c}/D$. However, a closer look reveals a small spread along the vertical and horizontal axes. The isosbestic exponent, $\alpha_{iso}$ reported here is located at the midpoint of this spread. We find $t_{\perp c}\approx 1.033\pm 0.0005$, $\alpha_{iso}=-0.35\pm0.01$ for $U/D = 1.75$ and $t_{\perp c}/D\approx 1.077\pm 0.001$, $\alpha_{iso}=-0.39\pm0.01$ for $U/D = 1.2$. 
Left(inset): A data collapse is achieved by a rescaling of the self-energy exponents shown in the main panel using the scaling ansatz, equation~\eqref{eq:scal}, reminiscent of scaling collapse in critical systems with a correlation length exponent $\nu=1.11\pm0.001$ for $U/D=1.75$ (also see Appendix~\ref{app2}).}
\label{fig:isos} 
\end{figure}

We find, as expected, that the exponent is non-integral in a wide region around the QCP. In Fig.~\ref{fig:isos} we plot this exponent $\alpha$ as a function of $t_\perp$ for different temperatures. As illustrated in Fig.~\ref{fig:isos}, the self-energy exponent ($\alpha$) changes smoothly across the QCP from a positive value to a negative value close to $-1$ with increasing $t_\perp$. Furthermore, if the power-law form of the self-energy at $t_{\perp c}$ is a genuine quantum effect then $\alpha$ should be independent of $T$ at this point; in other words the exponent at $t_{\perp c}$ should represent an isosbestic point in the $\alpha$ {\it vs.\ } $t_\perp$ plot as shown in Fig.\ref{fig:isos}. We label the isosbestic exponent as $\alpha_{iso}$, where the exponent is independent of temperature. .
Indeed, as demonstrated in Fig.\ref{fig:isos}, an isosbestic point, $\alpha_{iso}=-0.35\pm0.01$ is found at $t_\perp\approx (1.033\pm 0.0005)D$ for $U=1.75D$ (left panel) and $\alpha_{iso}=-0.39\pm0.01$ at $t_\perp\approx (1.077\pm 0.001)D$ for $U=1.2D$ (right panel). The error bar in $t_{\perp c}$ ($\alpha_{iso}$) indicate the spread along the horizontal (vertical) axis.

From the Matsubara quasiparticle fraction analysis (as shown in Fig.~\ref{fig:qpwvsT}) and the vanishing of the scales (Fig.~\ref{fig:energyscales}), we know that the two parameter sets, namely, $(U/D,t_\perp/D)=
(1.75,1.033)$ and $(1.2,1.077)$, represent quantum critical points.
Combining this inference with the self-energy analysis shown in Fig.~\ref{fig:isos}, yields the observation that the $t_\perp$ values corresponding to the quantum critical points are also isosbestic points. 
Indeed, it is well known that strongly correlated systems display {\it isosbestic} crossing points as a typical conspicuous feature in several observable quantities like the specific heat or the optical conductivity~\cite{isb1}. 
In quantum critical systems, the Binder cumulant obtained from local magnetization was shown to display such an intersection point that coincides with the QCP~\cite{glossop2011critical}. 
The identification of the QCP through an isosbestic point in the exponent that determines the Matsubara frequency dependence of the local, one-particle, self-energy is one of the highlights of this work.

As is well-known for isosbestic points, a rescaling of the axes should yield a data collapse\cite{pixley2013quantum}, and indeed, as seen in the inset of the left panel of Fig.~\ref{fig:isos}, the exponents for $U=1.75D$  collapse onto a universal curve for all temperatures, with the following scaling ansatz,
\begin{equation}
\alpha(t_\perp,T) = f_U\left(\frac{t_\perp -t_{\perp c}}{t_{\perp c} T^{1/\nu}}\right)\,,
\label{eq:scal}
\end{equation}
where $f_U$ is the scaling function.
A similar data collapse is observed for the exponents shown in the bottom right panel of Fig.~\ref{fig:scaleU1.2_and1.75} for $U=1.2D$ in Appendix~\ref{app2}. Furthermore, the scaling exponent $\nu$ that is typically associated with the correlation length exponent across a phase transition is found to be $\nu=1.11\pm0.001$ (using a procedure discussed in Appendix~\ref{app2}) for the parameter set, $U/D=1.75, t_{\perp c}\simeq 1.033D$ and $\nu=1.03\pm0.01$ for the second parameter set, $U=1.2D, t_{\perp c}\simeq 1.077$ . 
Perturbative renormalization group studies on the pseudogap Anderson and Kondo model predict $\nu=1$ for the pseudogap exponent, $r\ge 1$ when the solution is a level crossing transition and is therefore first-order. We will shortly prove that $r=-\alpha_{iso}$ for the model under investigation. In our case, we find  $\nu > 1$ and $0<r<1/2$ for which the transition is known to be second order\cite{Vojta_imp_QPT,ingersent2002critical}.  

The CTQMC method is constrained to work at finite temperatures, and becomes prohibitively expensive at lower temperatures. 
However, since the exponent $\alpha_{iso}$ at the quantum critical point is found to be an isosbestic, i.e independent of temperature in the investigated range, we conjecture that the exponent remains the same down to $T=0$, and such an extrapolation naturally leads to the inference that the self-energy has a power law form at the QCP, with a fractional exponent. 
In fact as Fig.~\ref{fig:isos} shows, as we traverse the $t_\perp$ axis, the $\alpha$ {\it vs.} $t_\perp$ becomes steeper with decreasing temperature, and extrapolating such behaviour down to $T=0$, the exponent should be $+1$ in the FL phase, changing abruptly to the isosbestic fractional exponent at the QCP, and then changing again immediately thereafter to $-1$ in the Mott insulating phase. 
This suggests that the low-frequency self-energy changes from having an analytic form ($\sim -i\omega_n$) in the FL phase to a branch point ($\sim -i\,{\rm sgn}(\omega_n)|\omega_n|^{\alpha_{iso}}$) at the QCP to a simple pole form ($1/i\omega_n$) in the Mott insulating phase. It is worth mentioning that further evidence of the interacting nature of the QCP could be provided via $\omega/T$ scaling of the one- and two-particle correlation functions evaluated on the real frequency axis. However, such analyses including the identification of the different critical exponents and their relation to the critical self-energy exponent $\alpha_{iso}$ will be taken up in later work.

Until now, we have confined ourselves to the imaginary frequency axis. In the next two sub-sections, we will use simple analytical arguments to deduce the manifestation of the power law form of the Matsubara frequency self-energy at the QCP, and density of states in the corresponding real-frequency quantities. 

\noindent
\subsection{Self-energy as a function of real frequency: behaviour at low frequencies} 
\label{subsec:soft_gap_dos}
The isosbestic exponent in the frequency dependence of the Matsubara frequency self-energy, must manifest in the real frequency dependence of the retarded self-energy. We show in this section that this is indeed the case. 
Using the arguments presented in the previous section, we may express the finite-temperature self-energy at the QCP as, $\Sigma(i\omega_n\to i0^+) = -iA\omega_{n}^{-r}$ where $r=-\alpha_{iso}$ and $\omega_n>0$. 
The corresponding real frequency self-energy may be obtained using analytic continuation, but here we employ the reverse route. Using an ansatz of the self-energy on the real-frequency axis, and employing the spectral representation, we derive the Matsubara frequency quantity, that matches the corresponding, numerically obtained quantity, thus validating the real-frequency ansatz.

We consider a power law form as an ansatz for $\rho_\Sigma$, namely,
\begin{align}
\rho_{\Sigma}(\omega)=-\frac{1}{\pi}{\rm Im}\Sigma(\omega)=B|\omega|^{-\gamma}, \quad B>0 \label{eq:SE_ansatz}, 
\end{align}
and ask if such a form, when substituted in the spectral representation, can yield the form, $\Sigma(i\omega_n\to i0^+) = -iA\omega_{n}^{-r}$ ($A>0$), thus avoiding analytic continuation. 
Indeed this is possible as we show below. 
Consequently, we would also know the relation between $\gamma$ and $r$. 

We begin with the spectral representation of the Matsubara frequency self-energy, specifically for the particle-hole symmetric case under consideration here as:
\begin{align}
    \Sigma(i\omega_n) = -2i\omega_n \int_{0}^{\infty}\frac{d\omega' \rho_{\Sigma}(\omega')}{\omega'^{2}+\omega_{n}^{2}} \label{eq:SE_spectral_rep}
\end{align}
 where $\displaystyle\rho_{\Sigma}(\omega)=(-1/\pi)\mathrm{Im}\Sigma(\omega)$, and $\Sigma(\omega)$ is the real frequency retarded self-energy.
We can substitute Equation~\eqref{eq:SE_ansatz} in Equation~\eqref{eq:SE_spectral_rep} to get (for $\omega_n>0$): 
\begin{align*}
    \Sigma(i\omega_n) 
    &=  -2iB\omega_n\int_{0}^{\Lambda}\frac{\omega'^{-\gamma} d\omega'}{\omega'^{2}+\omega_{n}^{2}} \\
   & = -2iB\omega_{n}^{-\gamma}\int_{0}^{\Lambda/\omega_n}\frac{dx}{x^{\gamma}(1+x^2)}\,,
\end{align*}
where $\Lambda$ is a UV-cutoff scale.
For $T\rightarrow 0$, $\Lambda/\omega_n\rightarrow \infty$ and since the integrand vanishes as $x^{2+\gamma}$ ($\gamma>0$) for large $x$, the upper limit of the integral can be considered to be $\infty$. Thus,
\begin{align}
    \Sigma(i\omega_n) = -2iB\,\text{sgn}(\omega_n)|\omega_{n}|^{-\gamma}\int_{0}^{\infty}\frac{dx}{x^{\gamma}(1+x^2)}.
    \label{eq:sigansatz}
\end{align}
From the numerically evaluated self-energy, we know that, at the isosbestic point,
\begin{align}
    \Sigma(i\omega_n\to 0) = -i \text{sgn}(\omega_n) A |\omega_{n}|^{-r} \label{eq:SE_Matsubara}\,.
\end{align}
Comparing Eq.~\ref{eq:sigansatz} with Eq.~\ref{eq:SE_Matsubara}, we get 
$\gamma = r$ and $\rho_{\Sigma}(\omega) = B|\omega|^{-r}$.  Now, given that the imaginary part of the self-energy has the form, $\rho_{\Sigma}(\omega) = B|\omega|^{-r}$, we can  derive the real part, ${\rm Re}\Sigma(\omega)$ through the Kramers-Kr\"onig transform: ${\rm Re}\Sigma(\omega) = -\mathcal{P}\int \,d\omega'\,\rho_{\Sigma}(\omega')(\omega-\omega')^{-1}$, and we get ${\rm Re}\Sigma(\omega)\simeq -B^\prime {\rm sgn}(\omega) |\omega|^{-r}$.
Combining the real and imaginary parts, we get the limiting low frequency self-energy as:
\begin{equation}
    \Sigma(\omega) = - (B^\prime{\rm sgn}(\omega) + i\pi B)|\omega|^{-r} =-\bar{B} |\omega|^{-r}\,, \label{eq:reself}
\end{equation}
with $B, B^\prime > 0$. In the following sub-section, we will show that the above form of the self-energy translates to a soft-gap form of the density of states.

Eq.~\ref{eq:SE_Matsubara} is also seen to be consistent with the results shown in Fig.~\ref{fig:qpwvsT}. To see this, we note that, for $n=0$,
$\Sigma(i\omega_0)=-i |\omega_0|^{-r}$, but since $\omega_0=\pi T$ and $Z_0(T)=(1-\Sigma(i\omega_0)/i\omega_0)^{-1}$, we get
$Z_0(T)=(1+|\omega_0|^{-(r+1)})$ which, in the low temperature limit, reduces to
$Z_0(T)\simeq T^{-(1+r)}$. As Fig.~\ref{fig:qpwvsT} shows, the exponent deduced from the power law is $1.36$, which is equal to $1-\alpha_{iso}$ deduced from the results of the left panel of Fig.~\ref{fig:isos}. A similar analysis has been carried out for $U=1.2D$, and we find $Z_0(T)\approx T^{-(1+r)}$, with $r=0.38\pm0.004$ which is consistent with the $r = -\alpha_{iso} = 0.39\pm0.01$ obtained for $U=1.2D$ from the right panel of Fig.~\ref{fig:isos}.
\subsection{Emergent soft-gap density of states at the QCP}
We can now use the real frequency power law form of the self-energy (equation~\eqref{eq:reself}), to derive the low frequency form of the $f$-electron Green's function on the real axis. Starting from the Green's function expression:
\begin{align}
    G_{f}(\omega) = \int d\epsilon \frac{\rho_0(\epsilon)}{z-\frac{V^2}{\omega^+-\epsilon-\frac{t_{\perp}^2}{{\omega^+-\epsilon}}}}.\label{eq:B7}
\end{align}
and using $\Sigma(\omega) = -\bar{B}|\omega|^{-r}$, with $z =\omega^+-\Sigma(\omega)\rightarrow\bar{B}|\omega|^{-r}$ as $\omega\rightarrow 0$, we get, to leading order,
\begin{align*}
    G_{f}(\omega)\approx& \int d\epsilon \frac{\rho_0(\epsilon)}{\bar{B}|\omega|^{-r}}=\frac{|\omega|^{r}}{\bar{B}}\,,
\end{align*}
by considering the most divergent term in the denominator. Thus, the spectral function of the f-density of states is given by:
\begin{align*}
\rho_{f}(\omega) = -\frac{1}{\pi}{\rm Im} G_f(\omega) = \frac{\mathrm{Im}\bar{B}/\pi}{(\mathrm{Re}\bar{B})^2+(\mathrm{Im}\bar{B})^2}|\omega|^{r}.
\end{align*}
The above expression shows that the $\rho_f(\omega)$ has a soft-gap form with the exponent $r$. The arguments presented in sections III.C and III.D  establish the link between the isosbestic exponent and the soft-gap exponent. Namely, if $\Sigma_{QCP}(i\omega_n\to 0)\approx -i\mathrm{sgn}(\omega_n)|\omega_{n}|^{-r}$, then $\rho_f(\omega)\approx|\omega|^{r}$. The isosbestic exponent and the soft-gap exponent of the spectrum have the same magnitude but the opposite sign. {\textbf {This emergent soft-gap density of states at the QCP represents our most important result.}} 

The inference of a soft-gap density of states can be further consolidated, by analytically continuing the CTQMC Greens function. Here we have used P\'ade approximation as implemented in Ref.~\cite{schott2016analytic} to get the real frequency spectra, $\rho_f(\omega)$. 
\begin{figure}[tbh!]

\includegraphics[clip=,scale=0.4]{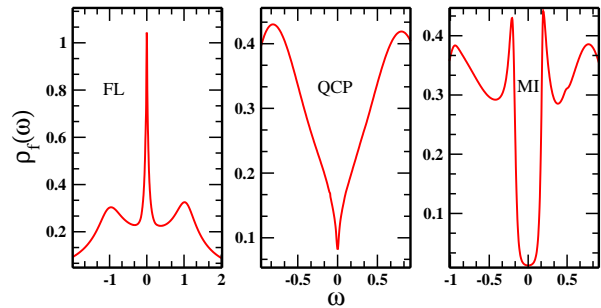}
\caption{The single particle density of states $\rho_f(\omega)$ obtained via analytic continuation of the Matsubara frequency Green's function $G_f(i\omega)$ using Pad\'e approximation, is shown for the Fermi liquid region (left panel) with $t_{\perp}/D = 0.8$, quantum critical point (middle panel) with $t_{\perp}/D = 1.033$ and the Mott insulator (right panel) with $t_{\perp}/D = 1.08$. The parameters used to obtain $G_f(i\omega)$ are, $U/D = 1.75$, $V/D = 0.44$. }
\label{fig:app_DOS} 
\end{figure}
The analytically continued $f$-spectra are shown in Fig.~\ref{fig:app_DOS} for the parameters $t_{\perp}=0.8$ in the left panel, $t_\perp=1.033=t_{\perp c}$ in the middle panel and $t_{\perp}=1.08$ in the right panel. In the left panel, $\rho_f(\omega)$ is peaked around $\omega=0$, which is the  Kondo resonance, characteristic of the FL phase. The MI phase (right panel)  exhibits a gapped $\rho_f(\omega$), representing a Mott insulator. 
Particularly at $t_\perp=1.033=t_{\perp c}$, the single particle density of states (DoS) $\rho_f(\omega; T=0.002)\sim |\omega|^r+a_0$ exhibits a power law, soft-gap DoS in agreement with the previous discussion. Thus, we have shown that, while the density of states has a Kondo resonance in the FL phase, and a gap in the MI phase, it has an emergent soft-gap form at the QCP.
\begin{figure}[htb!]
   
\includegraphics[clip=,width=1.0\columnwidth]{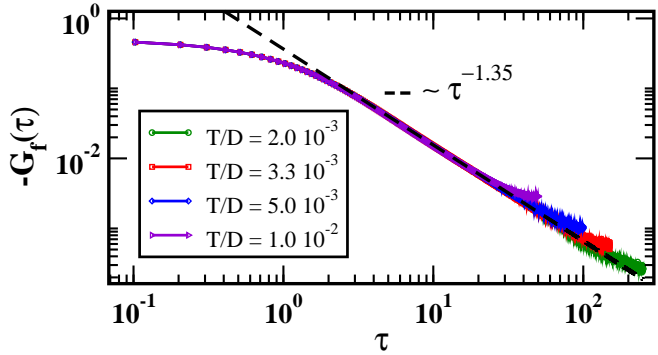}
\caption{The imaginary time Greens function $-G_f(\tau)$   as a function of the Matsubara time $\tau$ for $t_\perp = t_{\perp,c} = 1.033, U=1.75D$ and three different temperatures:(a) $T/D=0.01$ (purple), (b) $T/D=0.005$ (blue), (c) $T/D=0.003$ (red) d) $T/D = 0.002$ (green) at the QCP. The $\tau\gg 1$ behaviour is seen to fit very well to a power law ($\sim -\tau^{-(1+r)}$) (black dashed line) over more than a decade, and the extracted exponent, $r$, as obtained via this comparison is $r\approx0.35 \pm 0.01 \approx -\alpha_{iso}$~\cite{errorbar}, which is also independent of the temperature.}
  \label{fig:G_exponent} 
\end{figure}

Finally, it is straightforward to show that a soft-gap density of states of the form
$\rho_f(\omega) = |\omega|^r\theta(\Lambda - |\omega|)$, with
$\Lambda$ as a ultraviolet cutoff, results in a power law form in the Matsubara time dependence of the Green's function, namely
$G_f(\tau)\propto -\tau^{-(1+r)}$ for $\tau,\beta \gg 1$ (see Appendix~\ref{app4} for the derivation). 
Such a power law is indeed observed in the numerically obtained CTQMC results as shown in 
Fig.~\ref{fig:G_exponent} that shows $-G_f(\tau)$ as a function of $\tau$ as obtained for $U=1.75D$ and temperatures of $0.01$(purple), $0.005$(blue), $0.003$(red) and $0.002$(green). 
The $\tau\gg 1$ behaviour is clearly seen to be a power law, and the fit (dashed line) is seen to be valid over more than a decade for all the temperatures. 
The extracted exponent, $r$ from each of the fits is  $\sim 0.35\pm 0.01$, which is temperature independent and is seen to be the negative of the isosbestic exponent found from the self-energy analysis (Fig.~\ref{fig:isos}). 
Furthermore, as expected, the power law form does not fit the results obtained away from the QCP as shown in Fig.~\ref{fig:soft_gap_away} of Appendix \ref{app4}.



\subsection{The self-consistent hybridization function}
\label{subsec:hyb}
 Within DMFT, the lattice model is mapped onto a self-consistent impurity Anderson model, and our findings at the QCP indicate that a soft-gap Anderson model is emergent at the QCP. 
 We confirm this by showing that the imaginary part of the self-consistent hybridization is indeed a power law at the QCP. 
 \begin{figure}
    \centering
    
    \includegraphics[clip=,width=1.0\columnwidth]{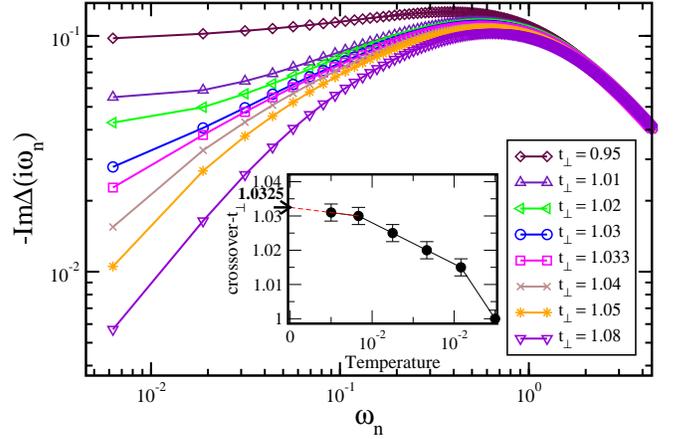}
    \caption{Main: The imaginary part of the self-consistent hybridization, $\Delta(i\omega_n)$, as a function of Matsubara frequency, for various $t_\perp$ values (in units of $D$) and $\beta=500$. Inset: The $t_\perp$ at which the imaginary part of the hybridization($-{\rm Im}\Delta(i\omega_n)$), at a given temperature, obeys a power-law at the maximum number of points is identified as the crossover point. The inset shows the crossover $t_\perp$ as a function of temperature.  The red dashed line shows a linear extrapolation to zero temperature, where the limiting value is seen to be the quantum critical value ($t_{\perp,c}$) within numerical resolution. The error bar represents the x-axis resolution to which the calculation is performed.}
    \label{fig:hyb}
\end{figure}
 To show this analytically, we need to revisit the arguments of section III.D, and compute sub-leading corrections to the Green's function for $T\to 0$ and $\omega_n\rightarrow 0^+$, and connect that to the DMFT hybridization function.

A detailed derivation is provided below. The main aim is to derive a relation between the isosbestic exponent($\alpha_{iso}$) and the hybridization exponent at the QCP. Given the Green's function,
\begin{align}
    G_f(i\omega_n) = \int \frac{d\epsilon \rho_0(\epsilon)}{\Omega_n-f_n(\epsilon)},
    \label{eq:gfiw}
\end{align}
where $\Omega_n = i\omega_n-\Sigma(i\omega_n)$ and
$f_n(\epsilon) = V^2\left[i\omega_n-\epsilon-t_{\perp}^2/(i\omega_n-\epsilon)\right]^{-1}$, we can substitute the low frequency form of $\Sigma(i\omega_n)$ to get
$\Omega_n\simeq i(\omega_n+A|\omega_{n}|^{-r})$ (with $r>0$ and  $r=-\alpha_{iso}$), which is divergent for $T\rightarrow 0$ or $\omega_n\rightarrow 0$. 
Hence, we can use the Taylor's expansion, to rewrite $G_f(i\omega_n)$ as
\begin{align}
  G_{f}^{-1}(i\omega_n)=\Omega_n - M_1+\frac{\left(M_{1}^2-M_2\right)}{\Omega_n} + {\mathcal{O}}\left(\frac{1}{\Omega_n^2}\right),
  \label{supeq:gf_expansion}
\end{align}
where the $k^{\rm th}$ moment is given by $M_k = \int d\epsilon \rho_0(\epsilon)f_n^k(\epsilon)$.
The self-consistent hybridization function is given by
\begin{equation}
\Delta(i\omega_n) =i\omega_n - \Sigma(i\omega_n) - G_f^{-1}(i\omega_n)\,,
\label{eq:schyb}
\end{equation}
and thus, to lowest order in $1/\Omega_n$,
the hybridization is given by $\Delta(i\omega_n) = M_1 - (M_{1}^2-M_2)/\Omega_n$.

For $\omega_n\rightarrow 0^+$, $M_1$ and $M_2$ are real constants, and it is easy to show that $M_1 = 0$, and $M_2$ is finite and positive definite(see appendix \ref{app5}). Hence, 
\begin{align}
    \Delta(\omega_n\to 0^+) =-\frac{iM_2|\omega_{n}|^{r}}{A}\,,
    \label{eq:hyb_exponent}
\end{align}
thus yielding a low-frequency power law form for the self-consistent hybridization function. We emphasize that, a divergent and dominant form of the self-energy and a vanishing $M_1=\int d\epsilon \rho_0(\epsilon) f_n(\epsilon)$ is important to ensure the validity of the moment expansion of the Green's function. Furthermore, the functional form of $f_n(\epsilon)$ in our model ensures that we will always have a vanishing $M_1$ and the low-frequency power-law form of the self-energy (Eq.~\eqref{eq:reself}) will always be the most singular function in the Hilbert transform. This is not true for other models like the standard periodic Anderson model~\cite{vojta_kircan_2003,logan2016mott,held2000mott}, where the self-energy vanishes as $\omega_n\to 0$.

We emphasise that the above analysis is correct only  in the zero temperature limit. 
In other words, a genuine power law vanishing (soft-gap) hybridization function is obtained at sufficiently low temperatures when $\omega_n\to 0$ makes true sense. At any non-zero temperature, the moments appearing in Eq.~\eqref{supeq:gf_expansion}, are dependent on $\omega_n$ (see Fig.~\ref{fig:m1m2} and discussion in Appendix(\ref{app5})) and contribute to the low-energy frequency behaviour of the hybridization function inducing deviations from a pristine power-law expected at the QCP, i.e. $t_\perp=t_{\perp c}$. 
Nonetheless, this frequency dependence becomes negligibly small as $T\to 0$, where $M_1\rightarrow 0$ and $M_2$ becomes purely real and positive-definite, producing a soft-gap hybridization function, $\Delta(i\omega_n\to 0) =-iA^{-1}M_2|\omega_{n}|^{r}$ as $T\to 0$ and $\omega_n\to 0$ (see Fig.~\ref{fig:10km1m2} in Appendix(\ref{app5}) for more details).

In Fig.~\ref{fig:hyb}, we plot the numerically obtained -Im$\Delta(i\omega_n)$ as a function of $\omega_n$ for different $t_\perp$'s and a fixed $U/D=1.75$ and $T/D=0.002$. 
The data in the main panel clearly demonstrate a change of curvature as the $t_\perp$ is tuned, similar to the Matsubara quasiparticle fraction (Fig.~\ref{fig:qpwvsT}) and the imaginary part of the self-energy (Fig.~\ref{fig:imsigma}), indicating a finite temperature crossover from a gapped to a metallic bath seen by the electrons as $t_\perp$ is decreased. 
Taking a closer look at the data suggests that this crossover occurs  at a value of $t_\perp$, that is not equal to the $t_{\perp c}$ at $\beta=500$. In fact, this crossover value is temperature-dependent, and only in the limit of $T\rightarrow 0$, it approaches the quantum critical value, $t_{\perp c}$, as the $T\rightarrow 0$ limiting value, represented by the red-dashed line in the inset of Fig.~\ref{fig:hyb} shows. The error bars for the crossover $t_\perp$ are estimated by the resolution at which the numerical calculations were performed. 

The finding of a power law form in the hybridization function at a $t_\perp\ne t_{\perp c}$ is thus seen to be a finite temperature effect, and as argued above, the impurity spectral function, the self-energy and the hybridization function acquire the same exponent only at $T\to 0$. We do not have a complete physical understanding of this finite temperature mismatch between hybridization and self-energy. We emphasize that this observation should not be interpreted as non-converged DMFT self-consistency since we have checked very carefully for convergence of the self-energy and the hybridization (please see the discussion in Appendix(\ref{app5} and Fig.~\ref{fig:exphyb}). Instead, it should be regarded as evidence that  temperatures as low as $T/D=10^{-4}$ need to be accessed to find the same exponent in the hybridization at $t_{\perp,c}$ within numerical resolution. However, we do not have the resources required to access such low temperatures at present.



We note that the hybridization as well as the density of states have a vanishing form as $T\to 0$, namely $\sim |\omega|^r$ ($r>0$)in the three-orbital model considered in this work, a behaviour that is quite distinct from the SGAM. 
In the latter, the hybridization has a vanishing form by choice for any temperature and, while the generalised FL phase has a divergent DoS, the local moment phase has a vanishing DoS, precisely as found here and in the critical Mott insulator in the single-band Hubbard model close to $U_{c1}^+$ ~\cite{eisenlohr2019mott}.  
We have also investigated a few other points
on the line of QCPs, and the soft-gap exponent appears to vary monotonically over the line of QCPs, in the $t_\perp$-$U$ plane, thus giving rise to a family of soft-gap Anderson models. A detailed investigation of the emergent family of the soft-gap Anderson models is underway and will be reported later. 
\noindent
\section{Outlook}
\label{sec:outlook}
Extensive investigations of quantum criticality in the soft-gap Anderson model (SGAM) over the last two decades have shown the QCP to be an interacting, non-Fermi liquid fixed point with a wide quantum critical scaling ($\omega/T$) region \cite{ingersent2002critical,glossop2005local,withoff1990phase,fritz2004phase,vojta2001kondo}. 
Such characteristic behaviour is inherited by the three-orbital model investigated here through the mapping, within DMFT, of the lattice Hamiltonian to the emergent SGAM wherein, low frequency, scale invariant power law excitations naturally emerge precisely at the QCP. 
This is reminiscent of a deconfined QCP~\cite{senthil2004deconfined}, where low energy fractional particles that are absent on either side of the QCP naturally emerge at the QCP.
The low frequency form of the self-energy having a branch point at the QCP, was also found in a dilaton gravity model using gauge-gravity correspondence \cite{iizuka2012holographic}, thus yielding a natural explanation of non-Fermi liquid behaviour emanating from the QCP. 
It would be interesting to explore how the emergent low energy critical modes interplay with long-range magnetic fluctuations~\cite{hu2017effects,si2001locally}, which would be predominantly relevant in lower dimensions. 
The particle-hole asymmetric SGAM with $r>1/2$ hosts a QCP between a Kondo screened and a local moment phase otherwise absent in the symmetric case. 
So a question remains as to the fate of the Mott QCP in the presence of particle-hole asymmetry. 
We anticipate that renormalization group approach\cite{mukherjee2020scaling1,mukherjee2020scaling2} may uncover the Hamiltonian flows of this three -orbital model, and hence lead to the underlying reasons for the emergence of the soft-gap Anderson models at quantum criticality. 

\acknowledgments
    NSV, VMK and SKK acknowledge funding from the Science and Engineering Research Board, India (EMR/2017/005398), National Supercomputing Mission, India (DST/NSM/R\&D\_HPC\_Applications/2021/26) and JNCASR. SKK acknowledges the DST-INSPIRE fellowship. We acknowledge discussions with Siddhartha Lal and Vladimir Dobr\"osavljevic. SS acknowledges from the Science and Engineering Research Board, India (SRG/2022/000495), (MTR/2022/000638) and IIT(ISM) Dhanbad (FRS(175)/2022-2023/PHYSICS).
\appendix
\section{Determination of Mott Gap}
\label{app1}

In order to extract the zero temperature Mott gap from the Matsubara Green's function obtained through CTQMC, we design a model density of states on the real frequency axis having certain unknown parameters, one of them being the $T=0$ Mott gap. The model DoS is then `Wick-rotated' onto the imaginary frequency axis, compared with the CTQMC Green's functions, and the Mott gap as well as the other parameters are obtained through a best fit. The procedure is detailed below.

\begin{figure}[htp]
\centering
\includegraphics[clip=,width=\columnwidth]{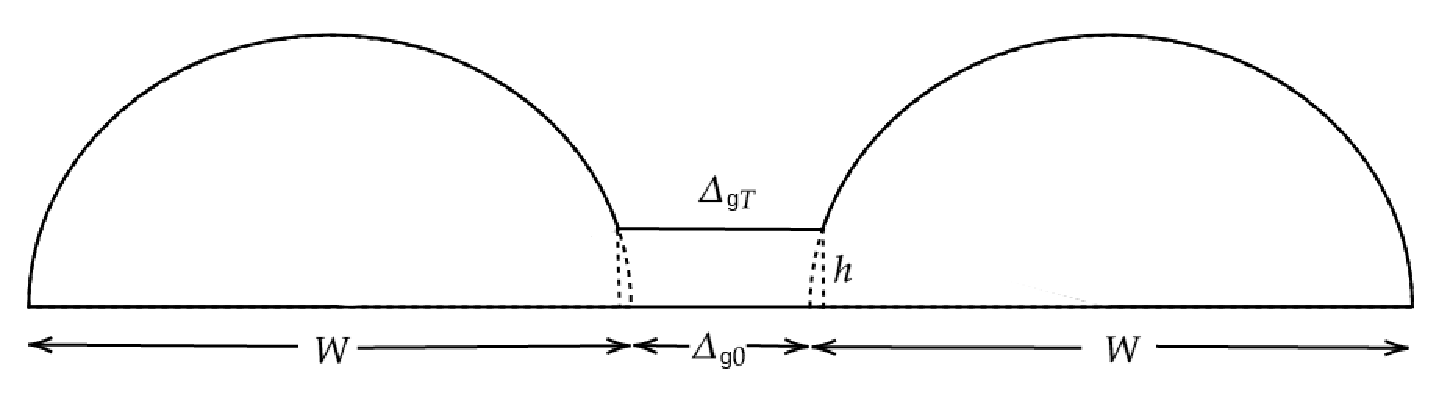}
\caption{Model DoS chosen to extract the Mott Gap $\Delta_{g0}$}
\label{fig:toy_ins_dos}
\end{figure}  

A model, gapped density of states (DoS) is shown in Fig. \ref{fig:toy_ins_dos} that has semi-elliptic bands flanking the gap. Such a model DoS is relevant for this work since our simulations consider a semi-elliptic 
DoS and therefore the Hubbard bands of the Mott insulator spectral function are semi-elliptic in form. Nonetheless, the strategy outlined should be applicable to any form of the Hubbard bands.
This model DoS is used as an ansatz for the putative real-frequency spectral function of the Mott insulator at a certain $t_\perp, U$ and temperature, $T$, with the $T=0$ Mott gap being denoted by $\Delta_{g0}$. At $T\ne 0$, we expect thermal filling of the Mott insulating gap. Let us denote the finite-temperature  `Mott gap' as $\Delta_{gT}$. 
Thus, at $T=0$, $\Delta_{gT}=\Delta_{g0}$. In order to mimic this finite temperature effect, we include a non-zero spectral weight at and around the Fermi level. Therefore, we modify the model spectral function by considering a parameter $h$ that mimics the Fermi level spectral weight. 
A mathematical representation for the the model DoS is the following:

\begin{align}
  \begin{split}
  \rho_{f}^{\rm{model}}(\epsilon)
  \begin{cases}
    & = h, \quad \forall\, |\epsilon|\leq \Delta_{gT}/2  \\
    & = \frac{4N}{\pi W}\left[1-\left(\frac{|\epsilon|-(W+\Delta_{g0})/2}{W/2}\right)^2\right]^{1/2}, \\
    &\quad \forall \,\, \Delta_{gT}/2\leq|\epsilon|\leq W+\Delta_{g0}/2
    \end{cases}
    \end{split}
    \label{eqn:toy_dos}
\end{align}
We note that the model DOS written above has five unknown parameters: $h, N, W, \Delta_{gT}, \Delta_{g0}$. However, there are three constraints using which we can reduce the number of unknowns to just two:
\begin{enumerate}[label=(\roman*)]
\item The finite temperature spectral density, $h$ may be obtained directly from CTQMC as $h=-G_f(\tau=\beta/2)/(\pi T)$. 
\item The constant $N$ may be determined in terms of the other parameters through the observation (from equation~\eqref{eqn:toy_dos} and Fig.~\ref{fig:toy_ins_dos}), that at $|\epsilon|=\frac{\Delta_{gT}}{2}$, $\rho_f^{\text{model}}(\epsilon)=h$ representing continuity of the model DoS:
\begin{equation}
N = \frac{h\pi W}{4 \cos\theta}
\label{eq:N}
\end{equation}
where $\sin\theta =1-(\Delta_{gT}-\Delta_{g0})/W$.
\item The spectral sum rule, i.e,
$\int\,d\epsilon\,\rho_f^{\rm model}(\epsilon)=1$, represents the third condition, and yields:
\begin{align}
    \frac{2N}{\pi}\left[\frac{\pi}{2}+\theta+\frac{\text{sin}(2\theta)}{2}\right]+h\Delta_{gT} = 1\label{eq:hN}\quad
\end{align}
\end{enumerate}

Using Eq.~\ref{eq:N} in Eq. \ref{eq:hN}, we get $W= 4\cos{\theta}\frac{(1/h)-\Delta_{gT}}{\pi+2\theta+\sin{(2\theta)}}$. Thus, we are left with only two fitting parameters, namely, $\theta$ and $\Delta_{gT}$. 

The strategy for obtaining the unknowns, i.e,
$\Delta_{gT}$ and $\theta$ is as follows:
\begin{enumerate}[label=(\roman*)]
\item Using the spectral representation, $G_{f}^{\rm{model}}(i\omega_n) = \int_{-\infty}^{\infty}d\epsilon \frac{\rho_{f}^{\rm{model}}(\epsilon)}{i\omega_n-\epsilon}$, the corresponding Matsubara Greens function may be calculated.
\item The model Green's function obtained in the first step above is then compared to the numerically obtained $G_f(i\omega_n)$ through CTQMC, to get the two unknown fitting parameters
through a best-fit procedure.
\end{enumerate}

As mentioned in step(i) above,
using the spectral representation, 
the model Matsubara Green's function may be obtained as:

\begin{widetext}
\begin{align}
    G_{f}^{\rm{model}}(i\omega_n; \Delta_{gT},W) &= -2ih\omega_n \left[\frac{W}{2\text{cos}\theta}\int_{-\theta}^{\pi/2}\frac{\text{cos}^2\bar{\theta}\, d\bar{\theta}}{\omega_{n}^2+\frac{W^2}{4}\left(\text{sin}\theta+\text{sin}\bar{\theta}+\frac{\Delta_{gT}}{W}\right)^2}+\frac{1}{\omega_n}\text{tan}^{-1}\left(\frac{\Delta_{gT}}{2\omega_n}\right)\right] \label{eq:modelgf}
\end{align}
\end{widetext}
Equation~\eqref{eq:modelgf} represents a two-parameter model Green's function, which may be used for fitting the imaginary part of the numerically obtained CTQMC Green's function, $G_f(i\omega_n)$, to get $\Delta_{gT}$ and $\theta$. For each combination of $\Delta_{gT}$ and $\theta$, the effective bandwidth is found through $W=4\cos\theta (1/h-\Delta_{gT})/(\pi+2\theta+\sin(2\theta))$.
The best fit then yields the zero temperature Mott gap, $\Delta_{g0}$ through $\sin\theta =1-(\Delta_{gT}-\Delta_{g0})/W$.

\begin{figure}[tbh]
   \centering
   
   \includegraphics[clip=,scale=0.4]{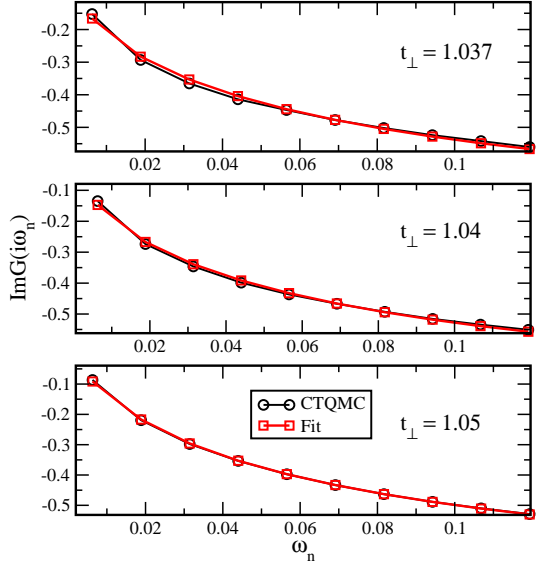}
   \caption{Matsubara Green's function is calculated using CTQMC for $T/D = 0.002$, and $t_{\perp}/D = 1.037 \text{ (Top panel)}, 1.04 \text{ (middle panel)}, 1.05 \text{ (bottom panel)}$, $U/D = 1.75$ and $V/D = 0.44$ and is fitted with the analytically calculated Matsubara Green's function in Eq.~\eqref{eq:modelgf} and hence the Mott gap ($\Delta_{gT}$) and effective bandwidth, $W$ are obtained.}
   \label{fig:app_fig2}
\end{figure}
Fig.~\ref{fig:app_fig2} demonstrates the imaginary part of the $f$-electron Matsubara Green's function calculated within DMFT-CTQMC fitted against the analytical form given in Equation~\eqref{eq:modelgf}, for $t_\perp=1.037$(top panel), $t_\perp=1.04$ (middle panel), $t_\perp=1.05$ (bottom panel). The agreement is seen to be excellent. 

\section{Determination of the Correlation-length exponent, $\nu$, through a scaling collapse of the isosbestic ($\alpha$ {\it vs.} $t_\perp$)}
\label{app2}

The $\alpha$ {\it vs.} $t_\perp$ curves shown in Fig.~\ref{fig:isos} display an isosbestic point at a certain $t_{\perp,c}$. If this point is to be identified as a quantum critical point, then all the curves passing through the isosbestic point must collapse onto a single curve when the $t_\perp$ is rescaled as 
\begin{equation}
X_\nu(\beta)=\beta^{1/\nu}\left(t_{\perp}/t_{\perp,c} - 1\right)\,.
\label{eq:trvar}
\end{equation}
However, the determination of $\nu$, the correlation length exponent, that determines the best scaling collapse, is not straightforward. The first step in this direction is to determine $\alpha$ as a function of $X_\nu$, given the known function $\alpha(t_\perp)$.
The mapping of $t_\perp$ to $X_\nu$ is linear, for a fixed $\beta$ and $\nu$, but the slope is temperature and $\nu$-dependent and care must be taken when trying to interpolate/extrapolate $\alpha(t_\perp)$ to $\alpha(X_\nu)$.

For any given $\nu$, when we rescale the $t_\perp$ axis, the window of the rescaled axis($X_\nu(t_\perp,\nu)$ axis) increases with an increase in $\beta$ for $\nu > 0$. Since the smallest window is obtained for high temperatures, in order to use only interpolation and avoid extrapolation of the data at lower temperatures, we consider $\beta_r = 100$ as the reference inverse temperature. For a given $\nu$, and $(t_\perp/t_{\perp,c}-1) \in[-\delta,\delta]$ window, we construct an $X_\nu\in\beta_r^{1/\nu}[-\delta,\delta]$ mesh with $N_x$ points  for the reference $\beta_r$.  Given the one-to-one correspondence between $t_\perp$ and $X_\nu$ for a given $\nu$ and $\beta$, it is straightforward to obtain $\alpha$ on the $X_\nu$ mesh. In practice, we calculate the corresponding self-energy exponent $\alpha(\beta,X_{\nu,i})$ using the Akima spline interpolation of the $\alpha$ {\it vs.} $t_\perp$ data. Note that a direct interpolation of $\alpha$ {\it vs.} $X_\nu$ is prone to error, hence should be avoided. We repeat this procedure for different $\beta$ values.

\begin{figure}[htp]
    \centering
   
    \includegraphics[clip=,scale=0.55]{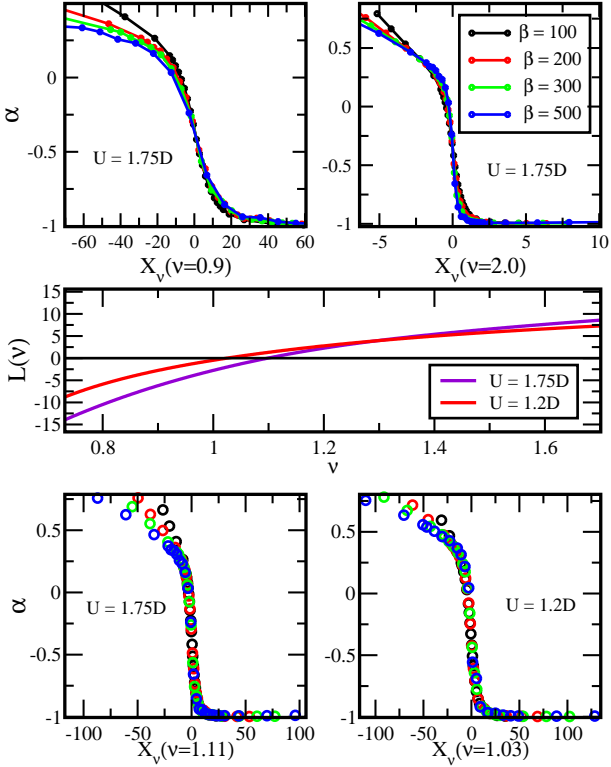}
    \caption{The top two panels show $\alpha$ {\it vs.} $X_\nu$ for $\nu=0.9$ (left) and $\nu=2.0$ (right) corresponding to $U=1.75D$. The middle panel shows the spread function, $L(\nu)$ as a function of $\nu$ for both $U=1.75D$ and $1.25D$. The $\nu$ that yields the zero crossing of the spread is identified as the correlation length exponent. The bottom two panels show the scaling collapse (of Fig.~\ref{fig:isos}) for $U = 1.75D$(Left) and $U = 1.2D$(Right) with the corresponding correlation length exponent $\nu(U = 1.75) = 1.11\pm0.001$ and $\nu(U = 1.2)=1.03\pm0.01$. The error in $\nu$ is found out from the corresponding error in determining the $t_{\perp,c}$ from the isosbestic plot in Fig. \ref{fig:isos}}.   
    \label{fig:scaleU1.2_and1.75}
\end{figure}

We do a visual inspection of $\alpha$ {\it vs.} $X_\nu$ for a range of $\nu$ values and observe that, within a chosen window around the QCP, as the $\nu$ is increased, the $\alpha$ {\it vs.} $X_\nu$ for $\beta=100$ and $\beta=500$ cross each other with the QCP being the pivot (see the top two panels of fig.~\ref{fig:scaleU1.2_and1.75}). This observation suggests that
in order to determine the scaling exponent that yields the best ``scaling collapse", we should look for a zero crossing of a spread function defined as
\begin{equation}
L(\nu) = \sum_{i=1}^{N_x}(2\theta(X_{\nu,i}) -1)\left(\alpha(100,X_{\nu,i})-\alpha(500,X_{\nu,i})\right)
\end{equation}
where $\theta(x)$ is a step-function. The finally obtained $\nu$ value corresponds to the root of $L(\nu)$, which also gives us the best scaling collapse around the QCP as shown in Fig.~\ref{fig:scaleU1.2_and1.75}. 

The choice of the window $[-\delta,\delta]$ around the QCP for computing the spread function is also made through the visual inspection of the crossings as shown in the top two panels of fig.~\ref{fig:scaleU1.2_and1.75}. For $U=1.75D$ and $1.2D$, we chose $\delta=0.133$ and $0.377$ respectively. The error associated in determination of $\nu$ arises from the error in the estimation of $t_{\perp c}$. In other words, we repeat the analysis for the maximum and minimum probable value of $t_{\perp c}$ and associate the respective spread in the $\nu$ obtained as the error.
We have verified that the exponent $\nu$ is quite robust with respect to a variation of $\delta$.

\section{Manifestation of a soft-gap density of states with exponent $r$ in the  Matsubara time dependence of the Green's function $G_f(\tau)$}
\label{app4}
\begin{figure}[tbh]
    \centering
  
    \includegraphics[clip=,scale=0.4]{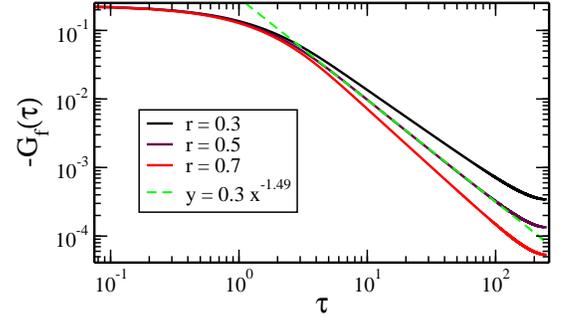}
    \caption{Numerically calculated $G_f(\tau)$ from the ansatz $\rho_f(\omega)=A |\omega|^r \Theta(D-|\omega|)$ with $D = 1.0 \,, \beta = 500.0$ using Eq.~\ref{supeq:gtau}. The $r$ is found to have its effect on $G_f(\tau)$ for $\tau\gg1$}
    \label{fig:fig14}
\end{figure}
In this section, we show that a real frequency, soft-gap DoS implies a power law dependence of the Matsubara Green's function, $G_f(\tau)$ on the imaginary time ($\tau$) axis. And the observation of power laws in the CTQMC derived  $G_f(\tau)$ may then be traced back to the existence of a real frequency soft-gap DoS through this proof, as done in Fig. \ref{fig:G_exponent} and the discussion associated with it.

Our starting point is a soft-gap density of states of the form, 
\begin{align}
 \rho_f(\omega)=A |\omega|^r \Theta(D-|\omega|)\,,
 \label{eq:spt}
\end{align}
where $|\omega|<D$, and $A$ may be found through normalization of the DoS as $A=(r+1)/2D^{r+1}$.
The Matsubara Greens function can be found by taking the Hilbert transform of the DoS (Eq.~\ref{eq:spt}) as:
\begin{equation}
    G_f(i\omega_n) = \int_{-\infty}^{\infty}A\frac{|\omega|^r}{i\omega_n-\omega}\Theta(D-|\omega|)d\omega\,.
    \label{eq:h}
\end{equation}
Fourier transforming the above, the Matsubara time Green's function, $G_f(\tau)$ can be represented in terms of $G_f(i\omega_n)$ as,
\begin{equation}
    G_f(\tau) = \frac{1}{\beta}\sum_{n}G_f(i\omega_n)e^{-i\omega_n\tau}
    \label{eq:f}
\end{equation}
Using Eqs.~\ref{eq:spt},\ref{eq:h} and \ref{eq:f},we get
\begin{align}
    &G_f(\tau) =-\int_{-D}^{0}d\omega \frac{(r+1)e^{-\omega \tau}|\omega|^r}{2D^{r+1}(1+e^{-\beta\omega})}&\label{supeq:gtau}\\
    &\quad\quad\,\,\,\,\qquad-\int_{0}^{D}d\omega \frac{(r+1)e^{-\omega \tau}|\omega|^r}{2D^{r+1}(1+e^{-\beta\omega})}\nonumber &\\
    &\quad\quad\,\,\,\,=\,\, G^{(1)} +G^{(2)}&\nonumber
\end{align}
\begin{figure}[tbh]
    \centering
  
    \includegraphics[clip=,scale=0.4]{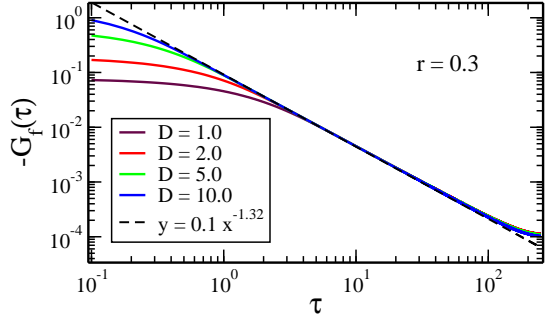}
    \caption{Numerically calculated $G_f(\tau)$ from the ansatz $\rho_f(\omega)=A |\omega|^r \Theta(D-|\omega|)$ with $r = 0.3$ and $\beta = 500.0$ using the Eq.~\eqref{supeq:gtau}. $G_f(\tau)\propto \tau^{-(1+r)}$ with $r \approx 0.3$, is independent of the cut-off $D$.}
    \label{fig:fig14b}
\end{figure}

Now, consider $G^{(1)}$ and taking the limit $\beta\rightarrow\infty$, we get,
\begin{align*}
    &G^{(1)} = -\int_{-D}^{0}d\omega \frac{(r+1)e^{-\omega \tau}|\omega|^r}{2D^{r+1}(1+e^{-\beta\omega})}& \\
    &\quad\quad\,\,\,\,\simeq \int_{0}^{D}d\omega \frac{(r+1)e^{-\beta\omega}e^{\omega \tau}\omega^r}{2D^{r+1}}&
\end{align*}
The above integral will have a contribution around $(2/\beta)^{r+1}$ for $\tau\rightarrow\beta/2$, and will vanish for $\beta\rightarrow\infty$, implying $G^{(1)} \rightarrow 0$ in the zero temperature limit. Now, taking $x = \omega\tau$ with $\beta\rightarrow\infty$ we get,
\begin{align}
G_f(\tau) & = G^{(2)} = -\tau^{-(r+1)} \int_{0}^{\infty} dx \frac{(r+1)e^{-x}|x|^r}{2D^{r+1}}\label{eq:gtau_integral}\\
 &\propto -\tau^{-(r+1)}\Gamma(r)\label{supeq:gtau_r}\,.
\end{align}
\begin{figure}[htp]
    \centering
    
    \includegraphics[clip=,scale=0.4]{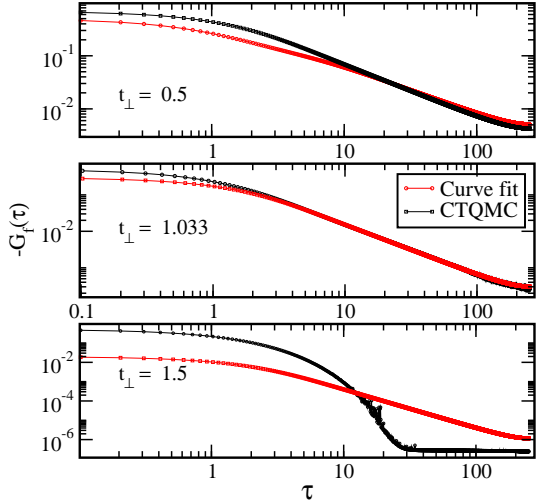}
    \caption{The Matsubara time Green's function $-G_f(\tau)$ obtained from CTQMC for $T/D = 2.0\times 10^{-3}$ is fitted to the integral Eq. \ref{supeq:gtau}. The top and the bottom panels represent the FL and the MI regimes where the integral equation does not fit the data. There is a clear fit only at the QCP ($t_{\perp} = 1.033$, middle panel). This indicates that the DoS away from the QCP does not have a power law form.}
    \label{fig:soft_gap_away}
\end{figure}
Thus, $G_f(\tau)$ depends on $r$ as given in the Eq. \ref{supeq:gtau_r}. We calculate $G_f(\tau)$ from Eq. \ref{supeq:gtau} directly by calculating the integral numerically without making the approximations involved in arriving at Eq.~\ref{supeq:gtau_r}. The results for varying $r$ and $D$ are plotted in Fig.~\ref{fig:fig14} and \ref{fig:fig14b}. The figures show that the exponent $r$ has its effect only for $\tau\gg1$, which is easily noticeable from Fig. \ref{fig:fig14}. We emphasise that these(Fig. \ref{fig:fig14} and Fig. \ref{fig:fig14b}) are not CTQMC results, but are obtained from the Eq. \ref{supeq:gtau}.

In Fig.~\ref{fig:soft_gap_away} we demonstrate sample fits of $G_f(\tau)$ obtained from CTQMC calculations with the power-law form, $\tau^{-(1+r)}$ for $t_\perp = 0.5$ (Kondo screened phase) and for $t_\perp = 1.5$ (Mott insulator phase) (top and bottom panel, respectively), where Eq.~\ref{supeq:gtau} does not fit the data at all, as should be expected far away from criticality. This is contrary to the middle panel of Fig.~\ref{fig:soft_gap_away}, representing the QCP, where the agreement is found to be the best, and a soft-gap power law form is seen over more than two decades (see Fig.~\ref{fig:G_exponent}).



\section{Detailed analysis on the hybridization function}
\label{app5}

\subsection{$T\rightarrow 0$ behaviour of the moments}

Now, let's try to understand the $T\to 0$ behaviour of $M_1$ and $M_2$ for $t_\perp = t_{\perp,\text{c}}$. We have,
\begin{align*}
    M_1 &= V^2\int_{-\infty}^{\infty}d\epsilon\frac{\rho_0(\epsilon)}{i\omega_n-\epsilon-\frac{t_{\perp,c}^2}{i\omega_n-\epsilon}}\\
    &= V^2\int_{-\infty}^{\infty}d\epsilon\,\rho_0(\epsilon)\frac{i\omega_n-\epsilon}{(i\omega_n-\epsilon)^2-t_{\perp,\text{c}}^2},\\
    &= \frac{V^2}{2}\int_{-\infty}^{\infty}d\epsilon\,\rho_0(\epsilon)\left(\frac{1}{i\omega_n-\epsilon-t_{\perp,\text{c}}}+\frac{1}{i\omega_n-\epsilon+t_{\perp,\text{c}}}\right),
\end{align*}
where, in the limit of $\omega_n\rightarrow0^+$, the real part of $M_1$ becomes (using $1/(x+i0^+)={\cal P}(1/x) -i\pi\delta(x)$),
\begin{align*}
    \text{Re}M_1 = \frac{V^2}{2}\int_{-\infty}^{\infty}d\epsilon\,\rho_0(\epsilon)\left[P\left(\frac{1}{-\epsilon-t_{\perp,\text{c}}}\right)+P\left(\frac{1}{-\epsilon+t_{\perp,\text{c}}}\right)\right],
\end{align*}
Since $\rho_0(\epsilon)\neq0 \;\forall \epsilon\in[-D,D]$ and $\rho_0(\epsilon)=0\; \forall |\epsilon| > D$, while $t_{\perp,\text{c}}>D$ for all the quantum critical points in the phase diagram, we get
\begin{align*}
    \text{Re}M_1 &= \frac{V^2}{2}\int_{-D}^{D}d\epsilon\,\rho_0(\epsilon)\left[\frac{1}{-\epsilon-t_{\perp,\text{c}}}+\frac{1}{-\epsilon+t_{\perp,\text{c}}}\right] \\
    &= \frac{V^2}{2}\int_{-D}^{D}d\epsilon\,\rho_0(\epsilon)\left[\frac{-2\epsilon}{\epsilon^2-t_{\perp}^2}\right] = 0, 
\end{align*}
for a p-h symmetric $\rho_0(\epsilon)$.
The imaginary part of $M_1$ becomes,
\begin{align*}
    \text{Im}M_1 = \frac{V^2}{2}\int_{-\infty}^{\infty}d\epsilon\,\rho_0(\epsilon)\left[-\pi\left(\delta(-\epsilon-t_{\perp,\text{c}})+\delta(-\epsilon+t_{\perp,\text{c}})\right)\right],
\end{align*}
With the same argument as we made earlier, since $|\epsilon|\leq D <t_{\perp,\text{c}}$ always, we get Im$M_1=0$. Thus, in the limit of $T\rightarrow 0$,
we have shown that $M_1\rightarrow 0$.

Now let us evaluate $M_2$, in the limit $i\omega_n\rightarrow i0^{+} = i\eta$ where $\eta\rightarrow 0^+$,
\begin{align*}
    M_2 &= \frac{V^4}{4}\int_{-\infty}^{\infty} d\epsilon\,\rho_0(\epsilon)\frac{(i\eta-\epsilon)^2}{[(i\eta-\epsilon)^2-t_{\perp,\text{c}}^2]^2}\\ 
    &= 
    \frac{V^4}{4}\int_{-\infty}^{\infty} d\epsilon\,\rho_0(\epsilon)\frac{\epsilon^2-2i\eta\epsilon}{(\epsilon^2-t_{\perp,\text{c}})^2-4i\eta\epsilon(\epsilon^2-t_{\perp,\text{c}}^2)}\\
    &= \frac{V^4}{4}\int_{-\infty}^\infty d\epsilon\,\rho_0(\epsilon)\epsilon^2\left[P\left(\frac{1}{(\epsilon^2-t_{\perp,\text{c}}^2)^2}\right)\right.\\
    &\qquad\qquad\left.+i4\pi\delta(\epsilon^2-t_{\perp,\text{c}}^2)\text{sgn}\left[\epsilon(\epsilon^2-t_{\perp,\text{c}}^2)\right]\right]
\end{align*}
From the same argument that we made earlier that $|\epsilon|<D$ and $t_{\perp,\text{c}}>D$, the imaginary part goes to zero. Hence only real part of $M_2$ survives,
\begin{align*}
    \text{Re}M_2 &= \frac{V^4}{4}\int_{-\infty}^{\infty}d\epsilon\,\rho_0(\epsilon)\epsilon^2P\left(\frac{1}{[\epsilon^2-t_{\perp,\text{c}}]^2}\right) \\
    &= \frac{V^4}{4}\int_{-D}^{D}d\epsilon\rho_0(\epsilon)\frac{\epsilon^2}{[\epsilon^2-t_{\perp,\text{c}}]^2}\,.
\end{align*}
From the above expression, it is easy to see that ${\rm Re} M_2$ is positive definite and finite, since the integrand is positive definite, and $t_{\perp c} > D$ for all the quantum critical points in the phase diagram. 

\subsection{Finite frequency behaviour of the moments}

\begin{figure}[H]
    \centering
    
    \includegraphics[clip=,scale=0.4]{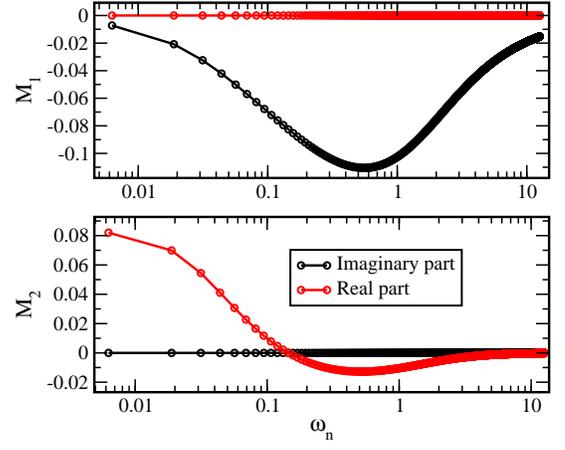}
    \caption{Plot of real and imaginary part of the moments $M_1$ and $M_2$ as a function of $\omega_n$ for $\beta=500$ and calculated at $t_\perp=t_{\perp,c}, V = 0.44D, U = 1.75D$.}
    \label{fig:m1m2}
\end{figure}

\begin{figure}[H]
    \centering
  
    \includegraphics[clip=,scale=0.4]{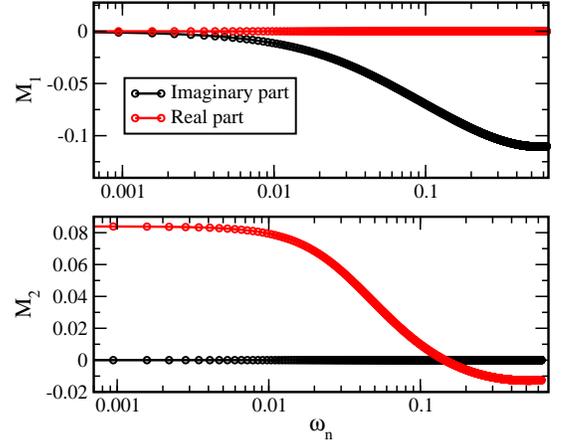}
    \caption{Plot of real and imaginary part of the moments $M_1$ and $M_2$ as a function of $\omega_n$ for $\beta=10000$ and calculated at $t_\perp=t_{\perp,c}, V = 0.44D, U = 1.75D$.}
    \label{fig:10km1m2}
\end{figure}

In order to understand the frequency behavior of the moments, $M_1$ and $M_2$, at low enough temperatures, we calculate the respective moments as a function of $\omega_n$ for two different inverse temperatures, namely, $\beta=500$ (Fig. \ref{fig:m1m2}) and $\beta=10^4$ (Fig. \ref{fig:10km1m2}), using a power law form for the self-energy. Please note that our CTQMC calculations are limited to $\beta \leq 500$, so the $\beta=10^4$ result shown in Fig.~\ref{fig:10km1m2} assumes that the same power law form of the self-energy holds at this temperature, and the moments are computed using the equation given below Eq.~\ref{supeq:gf_expansion}.

In Fig.~\ref{fig:m1m2}, we show the moments, $M_1$ and $M_2$ as a function of Matsubara frequency, as computed using the power law form of the self-energy at $\beta=500$. We see that as $\omega_n \rightarrow 0^+$, $M_1$ approaches a finite value at the lowest Matsubara frequency, which will result in a deviation from a perfect power law form of the hybridization function at the QCP. In order to consolidate this observation, we compute the moments at a much lower temperature, namely $\beta=10^4$, and as may be seen in Fig.~\ref{fig:10km1m2}, $M_1\rightarrow 0$. 

Fig.~\ref{fig:10km1m2} clearly demonstrates that $M_1\approx 0$ and $M_2$ is a real, positve definite constant in the limit of $T\to 0$ that naturally implies $\omega_n\to 0$, and, thereby, it is only in this limit that the hybridization function is power-law vanishing with the form, $\Delta(i\omega_n)\approx -\frac{iM_2|\omega_n|^r}{A}$, a genuine power-law. 
In fact, our analyses suggest that there would always exist a non-zero contribution from $M_1$ in the low-frequency behaviour of $\Delta(i\omega_n)$ at all non-zero temperatures. 
However, the low-frequency contribution will progressively tend to zero as we approach zero temperature. 
We infer from our analyses and emphasize that one may need to go as low as $\beta=10^4$ to demonstrate a true soft-gap hybridization at the QCP. Unfortunately, we do not have the resources required for accessing such low temperatures at present. 

Another question that we address now is that of DMFT self-consistency. From our analytical calculations (see Eq. \ref{eq:hyb_exponent}), we see that the exponent appearing in the hybridization is the same as that of the spectrum and the self-energy(see Eq. \ref{eq:SE_Matsubara}). However, the CTQMC results shown in Fig. \ref{fig:hyb} paint a different picture. The figure shows that the critical $t_\perp$ data does not have a clear power law form, but if we do somehow fit a power law form to the low-frequency data, we see an exponent of $0.42$, which differs from the self-energy exponent. Thus, a natural culprit for this mismatch may be identified as the absence of DMFT self-consistency. However, as we will show below, the real culprit is the limitation of our calculations to $\beta\leq 500$, due to the lack of resources, and not DMFT self-consistency.

\begin{figure}[H]
    \centering
    
    \includegraphics[clip=,scale=0.4]{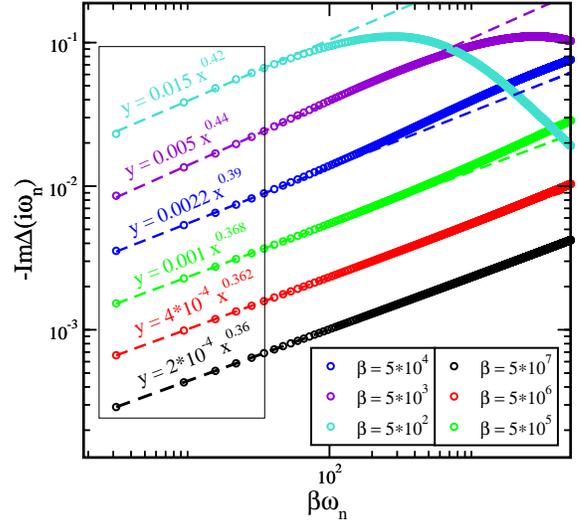}
    \caption{The imaginary part of the hybridization function, -Im$\Delta(i\omega_n)$, calculated at $t_\perp=1.033D,\, U=1.75D,\, V=0.44D$, is plotted as a function of the $\beta\omega_n$ for different inverse temperatures and is fitted with a power law. We assume a self-energy of the form, $\Sigma(i\omega_n)\sim iA|\omega_n|^{-r}$ with r = 0.36, as an ansatz in Equation~\eqref{eq:gfiw}(justified as the self-energy manifests an isosbestic point in temperature, at $t_\perp=1.033$). 
    The important thing to note is that (i) the range of the power law form extends to higher frequencies as we go down in temperature, and (ii) the value of the exponent approaches the value $r=0.36=-\alpha_{iso}$, 
    implying that a genuine soft-gap form with the expected soft-gap exponent, 
    $-\alpha_{\text{iso}}$, in the hybridization function emerges as $T \to 0$.}
    \label{fig:exphyb}
\end{figure}

In Fig.~\ref{fig:exphyb}, we show the imaginary part of the hybridization function as obtained through the DMFT Eqs. \ref{eq:gfiw}\ref{eq:schyb}, using a power law form of the self-energy $\Sigma(i\omega_n) = iA\omega_n^{-r}$ with $r=0.36$ is plotted as a function of $\beta\omega_n$. Note that this gives us the freedom to choose a desired temperature, but we again iterate that our CTQMC calculations are restricted to $\beta\leq500$. With this freedom, we obtain 
$-{\rm Im} \Delta(i\omega_n)$, for increasingly lower temperatures, and as seen in Fig.~\ref{fig:exphyb}, 
a power law form when fit to the low-frequency data, matches the numerical results over an increasingly larger frequency range as the temperature is lowered.
More importantly, the exponent of the power law decreases smoothly with decreasing temperature 
from $0.42$ to the isosbestic exponent, $0.36$.
This shows that the DMFT self-consistency, will not result in the same exponent in the hybridization at finite
temperatures. And only at zero temperature, a perfect power law form with exactly the same exponent will emerge in the hybridization at the quantum critical point.

\section{Additional numerical details}
\label{app6}
We use continuous-time quantum Monte-Carlo as implemented in TRIQS~\cite{seth2016triqs, werner2006continuous,werner2006hybridization,gull2008continuous,boehnke2011orthogonal,boehnke2015susceptibilities} open source package to study the finite temperature behaviour of the system. The Monte-Carlo parameters used in the calculations were:  n\_cycle = 200,000,  n\_warmup = 10,000, n\_length = 500. The parameter, n\_cycles, is increased to 1,000,000 for the final few DMFT iterations to reduce stochastic noise. Furthermore, close to the transition point we also increased the number of iterations for which n\_cycles = 1,000,000 to get the converged data. The number of Legendre polynomials(n\_l) considered are dependent on the temperature we choose. We used n\_l = 50 - 100 for inverse temperature ($\beta$) ranging from 50 to 500. The number of processors (Dual Intel Xeon Cascadelake 8268) employed was 48. For carrying out the analytic continuation through the Pad\'{e} approximation, we have used Beach's matrix formulation method, as implemented in the code\cite{schott2016analytic}, to find the Pad\'{e} approximants. The maximum number of Pad\'{e} approximant coefficients used for the analytic continuation is 80.

\bibliography{references}

\begin{thebibliography}{78}%
\makeatletter
\providecommand \@ifxundefined [1]{%
 \@ifx{#1\undefined}
}%
\providecommand \@ifnum [1]{%
 \ifnum #1\expandafter \@firstoftwo
 \else \expandafter \@secondoftwo
 \fi
}%
\providecommand \@ifx [1]{%
 \ifx #1\expandafter \@firstoftwo
 \else \expandafter \@secondoftwo
 \fi
}%
\providecommand \natexlab [1]{#1}%
\providecommand \enquote  [1]{``#1''}%
\providecommand \bibnamefont  [1]{#1}%
\providecommand \bibfnamefont [1]{#1}%
\providecommand \citenamefont [1]{#1}%
\providecommand \href@noop [0]{\@secondoftwo}%
\providecommand \href [0]{\begingroup \@sanitize@url \@href}%
\providecommand \@href[1]{\@@startlink{#1}\@@href}%
\providecommand \@@href[1]{\endgroup#1\@@endlink}%
\providecommand \@sanitize@url [0]{\catcode `\\12\catcode `\$12\catcode
  `\&12\catcode `\#12\catcode `\^12\catcode `\_12\catcode `\%12\relax}%
\providecommand \@@startlink[1]{}%
\providecommand \@@endlink[0]{}%
\providecommand \url  [0]{\begingroup\@sanitize@url \@url }%
\providecommand \@url [1]{\endgroup\@href {#1}{\urlprefix }}%
\providecommand \urlprefix  [0]{URL }%
\providecommand \Eprint [0]{\href }%
\providecommand \doibase [0]{http://dx.doi.org/}%
\providecommand \selectlanguage [0]{\@gobble}%
\providecommand \bibinfo  [0]{\@secondoftwo}%
\providecommand \bibfield  [0]{\@secondoftwo}%
\providecommand \translation [1]{[#1]}%
\providecommand \BibitemOpen [0]{}%
\providecommand \bibitemStop [0]{}%
\providecommand \bibitemNoStop [0]{.\EOS\space}%
\providecommand \EOS [0]{\spacefactor3000\relax}%
\providecommand \BibitemShut  [1]{\csname bibitem#1\endcsname}%
\let\auto@bib@innerbib\@empty
\bibitem [{\citenamefont {Coleman}\ and\ \citenamefont
  {Schofield}(2005)}]{coleman2005quantum}%
  \BibitemOpen
  \bibfield  {author} {\bibinfo {author} {\bibfnamefont {P.}~\bibnamefont
  {Coleman}}\ and\ \bibinfo {author} {\bibfnamefont {A.~J.}\ \bibnamefont
  {Schofield}},\ }\href@noop {} {\bibfield  {journal} {\bibinfo  {journal}
  {Nature}\ }\textbf {\bibinfo {volume} {433}},\ \bibinfo {pages} {226}
  (\bibinfo {year} {2005})}\BibitemShut {NoStop}%
\bibitem [{\citenamefont {Sachdev}\ and\ \citenamefont
  {Yin}(2010)}]{sachdev2010quantum}%
  \BibitemOpen
  \bibfield  {author} {\bibinfo {author} {\bibfnamefont {S.}~\bibnamefont
  {Sachdev}}\ and\ \bibinfo {author} {\bibfnamefont {X.}~\bibnamefont {Yin}},\
  }\href@noop {} {\bibfield  {journal} {\bibinfo  {journal} {Annals of
  Physics}\ }\textbf {\bibinfo {volume} {325}},\ \bibinfo {pages} {2} (\bibinfo
  {year} {2010})}\BibitemShut {NoStop}%
\bibitem [{\citenamefont {Vojta}(2006)}]{Vojta_imp_QPT}%
  \BibitemOpen
  \bibfield  {author} {\bibinfo {author} {\bibfnamefont {M.}~\bibnamefont
  {Vojta}},\ }\href {\doibase 10.1080/14786430500070396} {\bibfield  {journal}
  {\bibinfo  {journal} {Philosophical Magazine}\ }\textbf {\bibinfo {volume}
  {86}},\ \bibinfo {pages} {1807} (\bibinfo {year} {2006})},\ \Eprint
  {http://arxiv.org/abs/https://doi.org/10.1080/14786430500070396}
  {https://doi.org/10.1080/14786430500070396} \BibitemShut {NoStop}%
\bibitem [{\citenamefont {Vojta}(2003)}]{Vojta_2003}%
  \BibitemOpen
  \bibfield  {author} {\bibinfo {author} {\bibfnamefont {M.}~\bibnamefont
  {Vojta}},\ }\href {\doibase 10.1088/0034-4885/66/12/r01} {\bibfield
  {journal} {\bibinfo  {journal} {Reports on Progress in Physics}\ }\textbf
  {\bibinfo {volume} {66}},\ \bibinfo {pages} {2069} (\bibinfo {year}
  {2003})}\BibitemShut {NoStop}%
\bibitem [{\citenamefont {Si}\ and\ \citenamefont
  {Steglich}(2010)}]{Si_Steglich_2010}%
  \BibitemOpen
  \bibfield  {author} {\bibinfo {author} {\bibfnamefont {Q.}~\bibnamefont
  {Si}}\ and\ \bibinfo {author} {\bibfnamefont {F.}~\bibnamefont {Steglich}},\
  }\href {\doibase 10.1126/science.1191195} {\bibfield  {journal} {\bibinfo
  {journal} {Science}\ }\textbf {\bibinfo {volume} {329}},\ \bibinfo {pages}
  {1161} (\bibinfo {year} {2010})},\ \Eprint
  {http://arxiv.org/abs/https://www.science.org/doi/pdf/10.1126/science.1191195}
  {https://www.science.org/doi/pdf/10.1126/science.1191195} \BibitemShut
  {NoStop}%
\bibitem [{top(2016)}]{top_on_top}%
  \BibitemOpen
  \href {\doibase 10.1038/nphys3827} {\bibfield  {journal} {\bibinfo  {journal}
  {Nature Physics}\ }\textbf {\bibinfo {volume} {12}},\ \bibinfo {pages} {615}
  (\bibinfo {year} {2016})}\BibitemShut {NoStop}%
\bibitem [{\citenamefont {Continentino}(2017)}]{continentino2017topological}%
  \BibitemOpen
  \bibfield  {author} {\bibinfo {author} {\bibfnamefont {M.~A.}\ \bibnamefont
  {Continentino}},\ }\href@noop {} {\bibfield  {journal} {\bibinfo  {journal}
  {Physica B: Condensed Matter}\ }\textbf {\bibinfo {volume} {505}},\ \bibinfo
  {pages} {A1} (\bibinfo {year} {2017})}\BibitemShut {NoStop}%
\bibitem [{\citenamefont {Brink}\ \emph {et~al.}(2018)\citenamefont {Brink},
  \citenamefont {Gunn}, \citenamefont {Jose}, \citenamefont {Kosterlitz},\ and\
  \citenamefont {Phua}}]{brink2018topological}%
  \BibitemOpen
  \bibfield  {author} {\bibinfo {author} {\bibfnamefont {L.}~\bibnamefont
  {Brink}}, \bibinfo {author} {\bibfnamefont {M.}~\bibnamefont {Gunn}},
  \bibinfo {author} {\bibfnamefont {J.~V.}\ \bibnamefont {Jose}}, \bibinfo
  {author} {\bibfnamefont {J.~M.}\ \bibnamefont {Kosterlitz}}, \ and\ \bibinfo
  {author} {\bibfnamefont {K.~K.}\ \bibnamefont {Phua}},\ }\href@noop {} {\emph
  {\bibinfo {title} {Topological Phase Transitions and New Developments}}}\
  (\bibinfo  {publisher} {World Scientific},\ \bibinfo {year}
  {2018})\BibitemShut {NoStop}%
\bibitem [{\citenamefont {Mari{\'c}}\ \emph {et~al.}(2020)\citenamefont
  {Mari{\'c}}, \citenamefont {Giampaolo},\ and\ \citenamefont
  {Franchini}}]{maric2020quantum}%
  \BibitemOpen
  \bibfield  {author} {\bibinfo {author} {\bibfnamefont {V.}~\bibnamefont
  {Mari{\'c}}}, \bibinfo {author} {\bibfnamefont {S.~M.}\ \bibnamefont
  {Giampaolo}}, \ and\ \bibinfo {author} {\bibfnamefont {F.}~\bibnamefont
  {Franchini}},\ }\href@noop {} {\bibfield  {journal} {\bibinfo  {journal}
  {Communications Physics}\ }\textbf {\bibinfo {volume} {3}},\ \bibinfo {pages}
  {1} (\bibinfo {year} {2020})}\BibitemShut {NoStop}%
\bibitem [{\citenamefont {Gegenwart}\ \emph {et~al.}(2008)\citenamefont
  {Gegenwart}, \citenamefont {Si},\ and\ \citenamefont
  {Steglich}}]{gegenwart2008quantum}%
  \BibitemOpen
  \bibfield  {author} {\bibinfo {author} {\bibfnamefont {P.}~\bibnamefont
  {Gegenwart}}, \bibinfo {author} {\bibfnamefont {Q.}~\bibnamefont {Si}}, \
  and\ \bibinfo {author} {\bibfnamefont {F.}~\bibnamefont {Steglich}},\
  }\href@noop {} {\bibfield  {journal} {\bibinfo  {journal} {nature physics}\
  }\textbf {\bibinfo {volume} {4}},\ \bibinfo {pages} {186} (\bibinfo {year}
  {2008})}\BibitemShut {NoStop}%
\bibitem [{\citenamefont {Si}(2006)}]{si2006global}%
  \BibitemOpen
  \bibfield  {author} {\bibinfo {author} {\bibfnamefont {Q.}~\bibnamefont
  {Si}},\ }\href@noop {} {\bibfield  {journal} {\bibinfo  {journal} {Physica B:
  Condensed Matter}\ }\textbf {\bibinfo {volume} {378}},\ \bibinfo {pages} {23}
  (\bibinfo {year} {2006})}\BibitemShut {NoStop}%
\bibitem [{\citenamefont {Neumann}\ \emph {et~al.}(2007)\citenamefont
  {Neumann}, \citenamefont {Ny{\'e}ki}, \citenamefont {Cowan},\ and\
  \citenamefont {Saunders}}]{neumann2007bilayer}%
  \BibitemOpen
  \bibfield  {author} {\bibinfo {author} {\bibfnamefont {M.}~\bibnamefont
  {Neumann}}, \bibinfo {author} {\bibfnamefont {J.}~\bibnamefont {Ny{\'e}ki}},
  \bibinfo {author} {\bibfnamefont {B.}~\bibnamefont {Cowan}}, \ and\ \bibinfo
  {author} {\bibfnamefont {J.}~\bibnamefont {Saunders}},\ }\href@noop {}
  {\bibfield  {journal} {\bibinfo  {journal} {Science}\ }\textbf {\bibinfo
  {volume} {317}},\ \bibinfo {pages} {1356} (\bibinfo {year}
  {2007})}\BibitemShut {NoStop}%
\bibitem [{\citenamefont {Fritz}\ and\ \citenamefont
  {Vojta}(2013)}]{fritz2013physics}%
  \BibitemOpen
  \bibfield  {author} {\bibinfo {author} {\bibfnamefont {L.}~\bibnamefont
  {Fritz}}\ and\ \bibinfo {author} {\bibfnamefont {M.}~\bibnamefont {Vojta}},\
  }\href@noop {} {\bibfield  {journal} {\bibinfo  {journal} {Reports on
  Progress in Physics}\ }\textbf {\bibinfo {volume} {76}},\ \bibinfo {pages}
  {032501} (\bibinfo {year} {2013})}\BibitemShut {NoStop}%
\bibitem [{\citenamefont {Mitchell}\ and\ \citenamefont
  {Fritz}(2013)}]{mitchell2013kondo}%
  \BibitemOpen
  \bibfield  {author} {\bibinfo {author} {\bibfnamefont {A.~K.}\ \bibnamefont
  {Mitchell}}\ and\ \bibinfo {author} {\bibfnamefont {L.}~\bibnamefont
  {Fritz}},\ }\href@noop {} {\bibfield  {journal} {\bibinfo  {journal}
  {Physical Review B}\ }\textbf {\bibinfo {volume} {88}},\ \bibinfo {pages}
  {075104} (\bibinfo {year} {2013})}\BibitemShut {NoStop}%
\bibitem [{\citenamefont {Schneider}\ \emph {et~al.}(2011)\citenamefont
  {Schneider}, \citenamefont {Fritz}, \citenamefont {Anders}, \citenamefont
  {Benlagra},\ and\ \citenamefont {Vojta}}]{Anders-2-channel-SGAM}%
  \BibitemOpen
  \bibfield  {author} {\bibinfo {author} {\bibfnamefont {I.}~\bibnamefont
  {Schneider}}, \bibinfo {author} {\bibfnamefont {L.}~\bibnamefont {Fritz}},
  \bibinfo {author} {\bibfnamefont {F.~B.}\ \bibnamefont {Anders}}, \bibinfo
  {author} {\bibfnamefont {A.}~\bibnamefont {Benlagra}}, \ and\ \bibinfo
  {author} {\bibfnamefont {M.}~\bibnamefont {Vojta}},\ }\href {\doibase
  10.1103/PhysRevB.84.125139} {\bibfield  {journal} {\bibinfo  {journal} {Phys.
  Rev. B}\ }\textbf {\bibinfo {volume} {84}},\ \bibinfo {pages} {125139}
  (\bibinfo {year} {2011})}\BibitemShut {NoStop}%
\bibitem [{\citenamefont {Li}\ \emph {et~al.}(2017)\citenamefont {Li},
  \citenamefont {Jiang}, \citenamefont {Jian},\ and\ \citenamefont
  {Yao}}]{li2017fermion}%
  \BibitemOpen
  \bibfield  {author} {\bibinfo {author} {\bibfnamefont {Z.-X.}\ \bibnamefont
  {Li}}, \bibinfo {author} {\bibfnamefont {Y.-F.}\ \bibnamefont {Jiang}},
  \bibinfo {author} {\bibfnamefont {S.-K.}\ \bibnamefont {Jian}}, \ and\
  \bibinfo {author} {\bibfnamefont {H.}~\bibnamefont {Yao}},\ }\href@noop {}
  {\bibfield  {journal} {\bibinfo  {journal} {Nature communications}\ }\textbf
  {\bibinfo {volume} {8}},\ \bibinfo {pages} {1} (\bibinfo {year}
  {2017})}\BibitemShut {NoStop}%
\bibitem [{\citenamefont {Li}\ \emph {et~al.}(2020)\citenamefont {Li},
  \citenamefont {Li},\ and\ \citenamefont {Yao}}]{Meng1}%
  \BibitemOpen
  \bibfield  {author} {\bibinfo {author} {\bibfnamefont {B.-H.}\ \bibnamefont
  {Li}}, \bibinfo {author} {\bibfnamefont {Z.-X.}\ \bibnamefont {Li}}, \ and\
  \bibinfo {author} {\bibfnamefont {H.}~\bibnamefont {Yao}},\ }\href {\doibase
  10.1103/PhysRevB.101.085105} {\bibfield  {journal} {\bibinfo  {journal}
  {Phys. Rev. B}\ }\textbf {\bibinfo {volume} {101}},\ \bibinfo {pages}
  {085105} (\bibinfo {year} {2020})}\BibitemShut {NoStop}%
\bibitem [{\citenamefont {Zhou}\ \emph {et~al.}(2016)\citenamefont {Zhou},
  \citenamefont {Wang}, \citenamefont {Meng}, \citenamefont {Wang},\ and\
  \citenamefont {Wu}}]{Mott_MI_in_Dirac_sys_Meng2}%
  \BibitemOpen
  \bibfield  {author} {\bibinfo {author} {\bibfnamefont {Z.}~\bibnamefont
  {Zhou}}, \bibinfo {author} {\bibfnamefont {D.}~\bibnamefont {Wang}}, \bibinfo
  {author} {\bibfnamefont {Z.~Y.}\ \bibnamefont {Meng}}, \bibinfo {author}
  {\bibfnamefont {Y.}~\bibnamefont {Wang}}, \ and\ \bibinfo {author}
  {\bibfnamefont {C.}~\bibnamefont {Wu}},\ }\href {\doibase
  10.1103/PhysRevB.93.245157} {\bibfield  {journal} {\bibinfo  {journal} {Phys.
  Rev. B}\ }\textbf {\bibinfo {volume} {93}},\ \bibinfo {pages} {245157}
  (\bibinfo {year} {2016})}\BibitemShut {NoStop}%
\bibitem [{\citenamefont {Dabelow}\ \emph {et~al.}(2019)\citenamefont
  {Dabelow}, \citenamefont {Gies},\ and\ \citenamefont
  {Knorr}}]{FRG_Dirac_system}%
  \BibitemOpen
  \bibfield  {author} {\bibinfo {author} {\bibfnamefont {L.}~\bibnamefont
  {Dabelow}}, \bibinfo {author} {\bibfnamefont {H.}~\bibnamefont {Gies}}, \
  and\ \bibinfo {author} {\bibfnamefont {B.}~\bibnamefont {Knorr}},\ }\href
  {\doibase 10.1103/PhysRevD.99.125019} {\bibfield  {journal} {\bibinfo
  {journal} {Phys. Rev. D}\ }\textbf {\bibinfo {volume} {99}},\ \bibinfo
  {pages} {125019} (\bibinfo {year} {2019})}\BibitemShut {NoStop}%
\bibitem [{\citenamefont {Boyack}\ \emph {et~al.}(2021)\citenamefont {Boyack},
  \citenamefont {Yerzhakov},\ and\ \citenamefont
  {Maciejko}}]{boyack2021quantum}%
  \BibitemOpen
  \bibfield  {author} {\bibinfo {author} {\bibfnamefont {R.}~\bibnamefont
  {Boyack}}, \bibinfo {author} {\bibfnamefont {H.}~\bibnamefont {Yerzhakov}}, \
  and\ \bibinfo {author} {\bibfnamefont {J.}~\bibnamefont {Maciejko}},\
  }\href@noop {} {\bibfield  {journal} {\bibinfo  {journal} {The European
  Physical Journal Special Topics}\ }\textbf {\bibinfo {volume} {230}},\
  \bibinfo {pages} {979} (\bibinfo {year} {2021})}\BibitemShut {NoStop}%
\bibitem [{\citenamefont {Huang}\ \emph {et~al.}(2020)\citenamefont {Huang},
  \citenamefont {Guo}, \citenamefont {Maciejko}, \citenamefont {Scalettar},\
  and\ \citenamefont {Feng}}]{Scalettar_Dirac_fermion}%
  \BibitemOpen
  \bibfield  {author} {\bibinfo {author} {\bibfnamefont {Y.}~\bibnamefont
  {Huang}}, \bibinfo {author} {\bibfnamefont {H.}~\bibnamefont {Guo}}, \bibinfo
  {author} {\bibfnamefont {J.}~\bibnamefont {Maciejko}}, \bibinfo {author}
  {\bibfnamefont {R.~T.}\ \bibnamefont {Scalettar}}, \ and\ \bibinfo {author}
  {\bibfnamefont {S.}~\bibnamefont {Feng}},\ }\href {\doibase
  10.1103/PhysRevB.102.155152} {\bibfield  {journal} {\bibinfo  {journal}
  {Phys. Rev. B}\ }\textbf {\bibinfo {volume} {102}},\ \bibinfo {pages}
  {155152} (\bibinfo {year} {2020})}\BibitemShut {NoStop}%
\bibitem [{\citenamefont {Sato}\ \emph {et~al.}(2017)\citenamefont {Sato},
  \citenamefont {Hohenadler},\ and\ \citenamefont {Assaad}}]{Dirac_Assad}%
  \BibitemOpen
  \bibfield  {author} {\bibinfo {author} {\bibfnamefont {T.}~\bibnamefont
  {Sato}}, \bibinfo {author} {\bibfnamefont {M.}~\bibnamefont {Hohenadler}}, \
  and\ \bibinfo {author} {\bibfnamefont {F.~F.}\ \bibnamefont {Assaad}},\
  }\href {\doibase 10.1103/PhysRevLett.119.197203} {\bibfield  {journal}
  {\bibinfo  {journal} {Phys. Rev. Lett.}\ }\textbf {\bibinfo {volume} {119}},\
  \bibinfo {pages} {197203} (\bibinfo {year} {2017})}\BibitemShut {NoStop}%
\bibitem [{\citenamefont {Jian}\ and\ \citenamefont {Yao}(2017)}]{Weyl_QPT}%
  \BibitemOpen
  \bibfield  {author} {\bibinfo {author} {\bibfnamefont {S.-K.}\ \bibnamefont
  {Jian}}\ and\ \bibinfo {author} {\bibfnamefont {H.}~\bibnamefont {Yao}},\
  }\href {\doibase 10.1103/PhysRevB.96.155112} {\bibfield  {journal} {\bibinfo
  {journal} {Phys. Rev. B}\ }\textbf {\bibinfo {volume} {96}},\ \bibinfo
  {pages} {155112} (\bibinfo {year} {2017})}\BibitemShut {NoStop}%
\bibitem [{\citenamefont {Gonzalez-Buxton}\ and\ \citenamefont
  {Ingersent}(1996)}]{gonzalez1996stabilization}%
  \BibitemOpen
  \bibfield  {author} {\bibinfo {author} {\bibfnamefont {C.}~\bibnamefont
  {Gonzalez-Buxton}}\ and\ \bibinfo {author} {\bibfnamefont {K.}~\bibnamefont
  {Ingersent}},\ }\href@noop {} {\bibfield  {journal} {\bibinfo  {journal}
  {Physical Review B}\ }\textbf {\bibinfo {volume} {54}},\ \bibinfo {pages}
  {R15614} (\bibinfo {year} {1996})}\BibitemShut {NoStop}%
\bibitem [{\citenamefont {Chen}\ and\ \citenamefont
  {Jayaprakash}(1995)}]{chen1995kondo}%
  \BibitemOpen
  \bibfield  {author} {\bibinfo {author} {\bibfnamefont {K.}~\bibnamefont
  {Chen}}\ and\ \bibinfo {author} {\bibfnamefont {C.}~\bibnamefont
  {Jayaprakash}},\ }\href@noop {} {\bibfield  {journal} {\bibinfo  {journal}
  {Journal of Physics: Condensed Matter}\ }\textbf {\bibinfo {volume} {7}},\
  \bibinfo {pages} {L491} (\bibinfo {year} {1995})}\BibitemShut {NoStop}%
\bibitem [{\citenamefont {Ingersent}(1996)}]{ingersent1996behavior}%
  \BibitemOpen
  \bibfield  {author} {\bibinfo {author} {\bibfnamefont {K.}~\bibnamefont
  {Ingersent}},\ }\href@noop {} {\bibfield  {journal} {\bibinfo  {journal}
  {Physical Review B}\ }\textbf {\bibinfo {volume} {54}},\ \bibinfo {pages}
  {11936} (\bibinfo {year} {1996})}\BibitemShut {NoStop}%
\bibitem [{\citenamefont {Bulla}\ \emph {et~al.}(1997)\citenamefont {Bulla},
  \citenamefont {Pruschke},\ and\ \citenamefont {Hewson}}]{bulla1997anderson}%
  \BibitemOpen
  \bibfield  {author} {\bibinfo {author} {\bibfnamefont {R.}~\bibnamefont
  {Bulla}}, \bibinfo {author} {\bibfnamefont {T.}~\bibnamefont {Pruschke}}, \
  and\ \bibinfo {author} {\bibfnamefont {A.}~\bibnamefont {Hewson}},\
  }\href@noop {} {\bibfield  {journal} {\bibinfo  {journal} {Journal of
  Physics: Condensed Matter}\ }\textbf {\bibinfo {volume} {9}},\ \bibinfo
  {pages} {10463} (\bibinfo {year} {1997})}\BibitemShut {NoStop}%
\bibitem [{\citenamefont {Bulla}\ \emph {et~al.}(2000)\citenamefont {Bulla},
  \citenamefont {Glossop}, \citenamefont {Logan},\ and\ \citenamefont
  {Pruschke}}]{bulla2000soft}%
  \BibitemOpen
  \bibfield  {author} {\bibinfo {author} {\bibfnamefont {R.}~\bibnamefont
  {Bulla}}, \bibinfo {author} {\bibfnamefont {M.~T.}\ \bibnamefont {Glossop}},
  \bibinfo {author} {\bibfnamefont {D.~E.}\ \bibnamefont {Logan}}, \ and\
  \bibinfo {author} {\bibfnamefont {T.}~\bibnamefont {Pruschke}},\ }\href@noop
  {} {\bibfield  {journal} {\bibinfo  {journal} {Journal of Physics: Condensed
  Matter}\ }\textbf {\bibinfo {volume} {12}},\ \bibinfo {pages} {4899}
  (\bibinfo {year} {2000})}\BibitemShut {NoStop}%
\bibitem [{\citenamefont {Logan}\ and\ \citenamefont
  {Glossop}(2000)}]{logan2000local}%
  \BibitemOpen
  \bibfield  {author} {\bibinfo {author} {\bibfnamefont {D.~E.}\ \bibnamefont
  {Logan}}\ and\ \bibinfo {author} {\bibfnamefont {M.~T.}\ \bibnamefont
  {Glossop}},\ }\href@noop {} {\bibfield  {journal} {\bibinfo  {journal}
  {Journal of Physics: Condensed Matter}\ }\textbf {\bibinfo {volume} {12}},\
  \bibinfo {pages} {985} (\bibinfo {year} {2000})}\BibitemShut {NoStop}%
\bibitem [{\citenamefont {Ingersent}\ and\ \citenamefont
  {Si}(2002)}]{ingersent2002critical}%
  \BibitemOpen
  \bibfield  {author} {\bibinfo {author} {\bibfnamefont {K.}~\bibnamefont
  {Ingersent}}\ and\ \bibinfo {author} {\bibfnamefont {Q.}~\bibnamefont {Si}},\
  }\href@noop {} {\bibfield  {journal} {\bibinfo  {journal} {Physical review
  letters}\ }\textbf {\bibinfo {volume} {89}},\ \bibinfo {pages} {076403}
  (\bibinfo {year} {2002})}\BibitemShut {NoStop}%
\bibitem [{\citenamefont {Senthil}\ \emph {et~al.}(2004)\citenamefont
  {Senthil}, \citenamefont {Vishwanath}, \citenamefont {Balents}, \citenamefont
  {Sachdev},\ and\ \citenamefont {Fisher}}]{senthil2004deconfined}%
  \BibitemOpen
  \bibfield  {author} {\bibinfo {author} {\bibfnamefont {T.}~\bibnamefont
  {Senthil}}, \bibinfo {author} {\bibfnamefont {A.}~\bibnamefont {Vishwanath}},
  \bibinfo {author} {\bibfnamefont {L.}~\bibnamefont {Balents}}, \bibinfo
  {author} {\bibfnamefont {S.}~\bibnamefont {Sachdev}}, \ and\ \bibinfo
  {author} {\bibfnamefont {M.~P.}\ \bibnamefont {Fisher}},\ }\href@noop {}
  {\bibfield  {journal} {\bibinfo  {journal} {Science}\ }\textbf {\bibinfo
  {volume} {303}},\ \bibinfo {pages} {1490} (\bibinfo {year}
  {2004})}\BibitemShut {NoStop}%
\bibitem [{\citenamefont {Wang}\ \emph {et~al.}(2018)\citenamefont {Wang},
  \citenamefont {Okada}, \citenamefont {O’Neal}, \citenamefont {Zhou},
  \citenamefont {Walkup}, \citenamefont {Dhital}, \citenamefont {Hogan},
  \citenamefont {Clancy}, \citenamefont {Kim}, \citenamefont {Hu} \emph
  {et~al.}}]{wang2018disorder}%
  \BibitemOpen
  \bibfield  {author} {\bibinfo {author} {\bibfnamefont {Z.}~\bibnamefont
  {Wang}}, \bibinfo {author} {\bibfnamefont {Y.}~\bibnamefont {Okada}},
  \bibinfo {author} {\bibfnamefont {J.}~\bibnamefont {O’Neal}}, \bibinfo
  {author} {\bibfnamefont {W.}~\bibnamefont {Zhou}}, \bibinfo {author}
  {\bibfnamefont {D.}~\bibnamefont {Walkup}}, \bibinfo {author} {\bibfnamefont
  {C.}~\bibnamefont {Dhital}}, \bibinfo {author} {\bibfnamefont
  {T.}~\bibnamefont {Hogan}}, \bibinfo {author} {\bibfnamefont
  {P.}~\bibnamefont {Clancy}}, \bibinfo {author} {\bibfnamefont {Y.-J.}\
  \bibnamefont {Kim}}, \bibinfo {author} {\bibfnamefont {Y.}~\bibnamefont
  {Hu}},  \emph {et~al.},\ }\href@noop {} {\bibfield  {journal} {\bibinfo
  {journal} {Proceedings of the National Academy of Sciences}\ }\textbf
  {\bibinfo {volume} {115}},\ \bibinfo {pages} {11198} (\bibinfo {year}
  {2018})}\BibitemShut {NoStop}%
\bibitem [{\citenamefont {Misawa}\ and\ \citenamefont
  {Imada}(2007)}]{Imada_MIT_QCP}%
  \BibitemOpen
  \bibfield  {author} {\bibinfo {author} {\bibfnamefont {T.}~\bibnamefont
  {Misawa}}\ and\ \bibinfo {author} {\bibfnamefont {M.}~\bibnamefont {Imada}},\
  }\href {\doibase 10.1103/PhysRevB.75.115121} {\bibfield  {journal} {\bibinfo
  {journal} {Phys. Rev. B}\ }\textbf {\bibinfo {volume} {75}},\ \bibinfo
  {pages} {115121} (\bibinfo {year} {2007})}\BibitemShut {NoStop}%
\bibitem [{\citenamefont {Eisenlohr}\ \emph
  {et~al.}(2019{\natexlab{a}})\citenamefont {Eisenlohr}, \citenamefont {Lee},\
  and\ \citenamefont {Vojta}}]{Vojta_pseudogap_HM}%
  \BibitemOpen
  \bibfield  {author} {\bibinfo {author} {\bibfnamefont {H.}~\bibnamefont
  {Eisenlohr}}, \bibinfo {author} {\bibfnamefont {S.-S.~B.}\ \bibnamefont
  {Lee}}, \ and\ \bibinfo {author} {\bibfnamefont {M.}~\bibnamefont {Vojta}},\
  }\href {\doibase 10.1103/PhysRevB.100.155152} {\bibfield  {journal} {\bibinfo
   {journal} {Phys. Rev. B}\ }\textbf {\bibinfo {volume} {100}},\ \bibinfo
  {pages} {155152} (\bibinfo {year} {2019}{\natexlab{a}})}\BibitemShut
  {NoStop}%
\bibitem [{\citenamefont {Imada}\ \emph {et~al.}(1998)\citenamefont {Imada},
  \citenamefont {Fujimori},\ and\ \citenamefont {Tokura}}]{imada1998metal}%
  \BibitemOpen
  \bibfield  {author} {\bibinfo {author} {\bibfnamefont {M.}~\bibnamefont
  {Imada}}, \bibinfo {author} {\bibfnamefont {A.}~\bibnamefont {Fujimori}}, \
  and\ \bibinfo {author} {\bibfnamefont {Y.}~\bibnamefont {Tokura}},\
  }\href@noop {} {\bibfield  {journal} {\bibinfo  {journal} {Reviews of modern
  physics}\ }\textbf {\bibinfo {volume} {70}},\ \bibinfo {pages} {1039}
  (\bibinfo {year} {1998})}\BibitemShut {NoStop}%
\bibitem [{\citenamefont {Vu{\v{c}}i{\v{c}}evi{\'c}}\ \emph
  {et~al.}(2013)\citenamefont {Vu{\v{c}}i{\v{c}}evi{\'c}}, \citenamefont
  {Terletska}, \citenamefont {Tanaskovi{\'c}},\ and\ \citenamefont
  {Dobrosavljevi{\'c}}}]{vuvcivcevic2013finite}%
  \BibitemOpen
  \bibfield  {author} {\bibinfo {author} {\bibfnamefont {J.}~\bibnamefont
  {Vu{\v{c}}i{\v{c}}evi{\'c}}}, \bibinfo {author} {\bibfnamefont
  {H.}~\bibnamefont {Terletska}}, \bibinfo {author} {\bibfnamefont
  {D.}~\bibnamefont {Tanaskovi{\'c}}}, \ and\ \bibinfo {author} {\bibfnamefont
  {V.}~\bibnamefont {Dobrosavljevi{\'c}}},\ }\href@noop {} {\bibfield
  {journal} {\bibinfo  {journal} {Physical Review B}\ }\textbf {\bibinfo
  {volume} {88}},\ \bibinfo {pages} {075143} (\bibinfo {year}
  {2013})}\BibitemShut {NoStop}%
\bibitem [{\citenamefont {Li}\ \emph {et~al.}(2021)\citenamefont {Li},
  \citenamefont {Jiang}, \citenamefont {Li}, \citenamefont {Zhang},
  \citenamefont {Kang}, \citenamefont {Zhu}, \citenamefont {Watanabe},
  \citenamefont {Taniguchi}, \citenamefont {Chowdhury}, \citenamefont {Fu}
  \emph {et~al.}}]{li2021continuous}%
  \BibitemOpen
  \bibfield  {author} {\bibinfo {author} {\bibfnamefont {T.}~\bibnamefont
  {Li}}, \bibinfo {author} {\bibfnamefont {S.}~\bibnamefont {Jiang}}, \bibinfo
  {author} {\bibfnamefont {L.}~\bibnamefont {Li}}, \bibinfo {author}
  {\bibfnamefont {Y.}~\bibnamefont {Zhang}}, \bibinfo {author} {\bibfnamefont
  {K.}~\bibnamefont {Kang}}, \bibinfo {author} {\bibfnamefont {J.}~\bibnamefont
  {Zhu}}, \bibinfo {author} {\bibfnamefont {K.}~\bibnamefont {Watanabe}},
  \bibinfo {author} {\bibfnamefont {T.}~\bibnamefont {Taniguchi}}, \bibinfo
  {author} {\bibfnamefont {D.}~\bibnamefont {Chowdhury}}, \bibinfo {author}
  {\bibfnamefont {L.}~\bibnamefont {Fu}},  \emph {et~al.},\ }\href@noop {}
  {\bibfield  {journal} {\bibinfo  {journal} {Nature}\ }\textbf {\bibinfo
  {volume} {597}},\ \bibinfo {pages} {350} (\bibinfo {year}
  {2021})}\BibitemShut {NoStop}%
\bibitem [{\citenamefont {Sen}\ and\ \citenamefont
  {Vidhyadhiraja}(2016)}]{sen2016quantum}%
  \BibitemOpen
  \bibfield  {author} {\bibinfo {author} {\bibfnamefont {S.}~\bibnamefont
  {Sen}}\ and\ \bibinfo {author} {\bibfnamefont {N.}~\bibnamefont
  {Vidhyadhiraja}},\ }\href@noop {} {\bibfield  {journal} {\bibinfo  {journal}
  {Physical Review B}\ }\textbf {\bibinfo {volume} {93}},\ \bibinfo {pages}
  {155136} (\bibinfo {year} {2016})}\BibitemShut {NoStop}%
\bibitem [{\citenamefont {Hu}\ \emph {et~al.}(2017)\citenamefont {Hu},
  \citenamefont {Scalettar}, \citenamefont {Huang},\ and\ \citenamefont
  {Moritz}}]{hu2017effects}%
  \BibitemOpen
  \bibfield  {author} {\bibinfo {author} {\bibfnamefont {W.}~\bibnamefont
  {Hu}}, \bibinfo {author} {\bibfnamefont {R.~T.}\ \bibnamefont {Scalettar}},
  \bibinfo {author} {\bibfnamefont {E.~W.}\ \bibnamefont {Huang}}, \ and\
  \bibinfo {author} {\bibfnamefont {B.}~\bibnamefont {Moritz}},\ }\href@noop {}
  {\bibfield  {journal} {\bibinfo  {journal} {Physical Review B}\ }\textbf
  {\bibinfo {volume} {95}},\ \bibinfo {pages} {235122} (\bibinfo {year}
  {2017})}\BibitemShut {NoStop}%
\bibitem [{\citenamefont {dos Anjos Sousa-J{\'u}nior}\ \emph
  {et~al.}(2020)\citenamefont {dos Anjos Sousa-J{\'u}nior}, \citenamefont
  {de~Lima}, \citenamefont {Costa},\ and\ \citenamefont {dos
  Santos}}]{dos2020superconducting}%
  \BibitemOpen
  \bibfield  {author} {\bibinfo {author} {\bibfnamefont {S.}~\bibnamefont {dos
  Anjos Sousa-J{\'u}nior}}, \bibinfo {author} {\bibfnamefont {J.~P.}\
  \bibnamefont {de~Lima}}, \bibinfo {author} {\bibfnamefont {N.~C.}\
  \bibnamefont {Costa}}, \ and\ \bibinfo {author} {\bibfnamefont {R.~R.}\
  \bibnamefont {dos Santos}},\ }\href@noop {} {\bibfield  {journal} {\bibinfo
  {journal} {Physical Review Research}\ }\textbf {\bibinfo {volume} {2}},\
  \bibinfo {pages} {033168} (\bibinfo {year} {2020})}\BibitemShut {NoStop}%
\bibitem [{\citenamefont {Yang}\ \emph {et~al.}(2019)\citenamefont {Yang},
  \citenamefont {Wang}, \citenamefont {Ma},\ and\ \citenamefont
  {Chen}}]{yang2019degenerate}%
  \BibitemOpen
  \bibfield  {author} {\bibinfo {author} {\bibfnamefont {J.}~\bibnamefont
  {Yang}}, \bibinfo {author} {\bibfnamefont {Q.}~\bibnamefont {Wang}}, \bibinfo
  {author} {\bibfnamefont {T.}~\bibnamefont {Ma}}, \ and\ \bibinfo {author}
  {\bibfnamefont {Q.}~\bibnamefont {Chen}},\ }\href@noop {} {\bibfield
  {journal} {\bibinfo  {journal} {Physical Review B}\ }\textbf {\bibinfo
  {volume} {99}},\ \bibinfo {pages} {245136} (\bibinfo {year}
  {2019})}\BibitemShut {NoStop}%
\bibitem [{\citenamefont {Georges}\ \emph {et~al.}(1996)\citenamefont
  {Georges}, \citenamefont {Kotliar}, \citenamefont {Krauth},\ and\
  \citenamefont {Rozenberg}}]{dmft_rmp}%
  \BibitemOpen
  \bibfield  {author} {\bibinfo {author} {\bibfnamefont {A.}~\bibnamefont
  {Georges}}, \bibinfo {author} {\bibfnamefont {G.}~\bibnamefont {Kotliar}},
  \bibinfo {author} {\bibfnamefont {W.}~\bibnamefont {Krauth}}, \ and\ \bibinfo
  {author} {\bibfnamefont {M.~J.}\ \bibnamefont {Rozenberg}},\ }\href {\doibase
  10.1103/RevModPhys.68.13} {\bibfield  {journal} {\bibinfo  {journal} {Rev.
  Mod. Phys.}\ }\textbf {\bibinfo {volume} {68}},\ \bibinfo {pages} {13}
  (\bibinfo {year} {1996})}\BibitemShut {NoStop}%
\bibitem [{\citenamefont {Seth}\ \emph {et~al.}(2016)\citenamefont {Seth},
  \citenamefont {Krivenko}, \citenamefont {Ferrero},\ and\ \citenamefont
  {Parcollet}}]{seth2016triqs}%
  \BibitemOpen
  \bibfield  {author} {\bibinfo {author} {\bibfnamefont {P.}~\bibnamefont
  {Seth}}, \bibinfo {author} {\bibfnamefont {I.}~\bibnamefont {Krivenko}},
  \bibinfo {author} {\bibfnamefont {M.}~\bibnamefont {Ferrero}}, \ and\
  \bibinfo {author} {\bibfnamefont {O.}~\bibnamefont {Parcollet}},\ }\href@noop
  {} {\bibfield  {journal} {\bibinfo  {journal} {Computer Physics
  Communications}\ }\textbf {\bibinfo {volume} {200}},\ \bibinfo {pages} {274}
  (\bibinfo {year} {2016})}\BibitemShut {NoStop}%
\bibitem [{\citenamefont {Werner}\ \emph {et~al.}(2006)\citenamefont {Werner},
  \citenamefont {Comanac}, \citenamefont {De’Medici}, \citenamefont
  {Troyer},\ and\ \citenamefont {Millis}}]{werner2006continuous}%
  \BibitemOpen
  \bibfield  {author} {\bibinfo {author} {\bibfnamefont {P.}~\bibnamefont
  {Werner}}, \bibinfo {author} {\bibfnamefont {A.}~\bibnamefont {Comanac}},
  \bibinfo {author} {\bibfnamefont {L.}~\bibnamefont {De’Medici}}, \bibinfo
  {author} {\bibfnamefont {M.}~\bibnamefont {Troyer}}, \ and\ \bibinfo {author}
  {\bibfnamefont {A.~J.}\ \bibnamefont {Millis}},\ }\href@noop {} {\bibfield
  {journal} {\bibinfo  {journal} {Physical Review Letters}\ }\textbf {\bibinfo
  {volume} {97}},\ \bibinfo {pages} {076405} (\bibinfo {year}
  {2006})}\BibitemShut {NoStop}%
\bibitem [{\citenamefont {Werner}\ and\ \citenamefont
  {Millis}(2006)}]{werner2006hybridization}%
  \BibitemOpen
  \bibfield  {author} {\bibinfo {author} {\bibfnamefont {P.}~\bibnamefont
  {Werner}}\ and\ \bibinfo {author} {\bibfnamefont {A.~J.}\ \bibnamefont
  {Millis}},\ }\href@noop {} {\bibfield  {journal} {\bibinfo  {journal}
  {Physical Review B}\ }\textbf {\bibinfo {volume} {74}},\ \bibinfo {pages}
  {155107} (\bibinfo {year} {2006})}\BibitemShut {NoStop}%
\bibitem [{\citenamefont {Gull}(2008)}]{gull2008continuous}%
  \BibitemOpen
  \bibfield  {author} {\bibinfo {author} {\bibfnamefont {E.}~\bibnamefont
  {Gull}},\ }\emph {\bibinfo {title} {Continuous-time quantum Monte Carlo
  algorithms for fermions}},\ \href@noop {} {Ph.D. thesis},\ \bibinfo  {school}
  {ETH Zurich} (\bibinfo {year} {2008})\BibitemShut {NoStop}%
\bibitem [{\citenamefont {Boehnke}\ \emph {et~al.}(2011)\citenamefont
  {Boehnke}, \citenamefont {Hafermann}, \citenamefont {Ferrero}, \citenamefont
  {Lechermann},\ and\ \citenamefont {Parcollet}}]{boehnke2011orthogonal}%
  \BibitemOpen
  \bibfield  {author} {\bibinfo {author} {\bibfnamefont {L.}~\bibnamefont
  {Boehnke}}, \bibinfo {author} {\bibfnamefont {H.}~\bibnamefont {Hafermann}},
  \bibinfo {author} {\bibfnamefont {M.}~\bibnamefont {Ferrero}}, \bibinfo
  {author} {\bibfnamefont {F.}~\bibnamefont {Lechermann}}, \ and\ \bibinfo
  {author} {\bibfnamefont {O.}~\bibnamefont {Parcollet}},\ }\href@noop {}
  {\bibfield  {journal} {\bibinfo  {journal} {Physical Review B}\ }\textbf
  {\bibinfo {volume} {84}},\ \bibinfo {pages} {075145} (\bibinfo {year}
  {2011})}\BibitemShut {NoStop}%
\bibitem [{\citenamefont {Boehnke}(2015)}]{boehnke2015susceptibilities}%
  \BibitemOpen
  \bibfield  {author} {\bibinfo {author} {\bibfnamefont {L.~V.}\ \bibnamefont
  {Boehnke}},\ }\emph {\bibinfo {title} {Susceptibilities in materials with
  multiple strongly correlated orbitals}},\ \href@noop {} {Ph.D. thesis},\
  \bibinfo  {school} {Staats-und Universit{\"a}tsbibliothek Hamburg Carl von
  Ossietzky} (\bibinfo {year} {2015})\BibitemShut {NoStop}%
\bibitem [{\citenamefont {Gull}\ \emph {et~al.}(2011)\citenamefont {Gull},
  \citenamefont {Millis}, \citenamefont {Lichtenstein}, \citenamefont
  {Rubtsov}, \citenamefont {Troyer},\ and\ \citenamefont
  {Werner}}]{CTQMC_review}%
  \BibitemOpen
  \bibfield  {author} {\bibinfo {author} {\bibfnamefont {E.}~\bibnamefont
  {Gull}}, \bibinfo {author} {\bibfnamefont {A.~J.}\ \bibnamefont {Millis}},
  \bibinfo {author} {\bibfnamefont {A.~I.}\ \bibnamefont {Lichtenstein}},
  \bibinfo {author} {\bibfnamefont {A.~N.}\ \bibnamefont {Rubtsov}}, \bibinfo
  {author} {\bibfnamefont {M.}~\bibnamefont {Troyer}}, \ and\ \bibinfo {author}
  {\bibfnamefont {P.}~\bibnamefont {Werner}},\ }\href@noop {} {\bibfield
  {journal} {\bibinfo  {journal} {Reviews of Modern Physics}\ }\textbf
  {\bibinfo {volume} {83}},\ \bibinfo {pages} {349} (\bibinfo {year}
  {2011})}\BibitemShut {NoStop}%
\bibitem [{\citenamefont {Peters}\ \emph {et~al.}(2006)\citenamefont {Peters},
  \citenamefont {Pruschke},\ and\ \citenamefont
  {Anders}}]{peters2006numerical}%
  \BibitemOpen
  \bibfield  {author} {\bibinfo {author} {\bibfnamefont {R.}~\bibnamefont
  {Peters}}, \bibinfo {author} {\bibfnamefont {T.}~\bibnamefont {Pruschke}}, \
  and\ \bibinfo {author} {\bibfnamefont {F.~B.}\ \bibnamefont {Anders}},\
  }\href@noop {} {\bibfield  {journal} {\bibinfo  {journal} {Physical Review
  B}\ }\textbf {\bibinfo {volume} {74}},\ \bibinfo {pages} {245114} (\bibinfo
  {year} {2006})}\BibitemShut {NoStop}%
\bibitem [{\citenamefont {Jarrell}\ \emph {et~al.}(1993)\citenamefont
  {Jarrell}, \citenamefont {Akhlaghpour},\ and\ \citenamefont
  {Pruschke}}]{jarrell1993periodic}%
  \BibitemOpen
  \bibfield  {author} {\bibinfo {author} {\bibfnamefont {M.}~\bibnamefont
  {Jarrell}}, \bibinfo {author} {\bibfnamefont {H.}~\bibnamefont
  {Akhlaghpour}}, \ and\ \bibinfo {author} {\bibfnamefont {T.}~\bibnamefont
  {Pruschke}},\ }\href@noop {} {\bibfield  {journal} {\bibinfo  {journal}
  {Physical review letters}\ }\textbf {\bibinfo {volume} {70}},\ \bibinfo
  {pages} {1670} (\bibinfo {year} {1993})}\BibitemShut {NoStop}%
\bibitem [{\citenamefont {Vidhyadhiraja}\ and\ \citenamefont
  {Logan}(2004)}]{vidhyadhiraja2004dynamics}%
  \BibitemOpen
  \bibfield  {author} {\bibinfo {author} {\bibfnamefont {N.}~\bibnamefont
  {Vidhyadhiraja}}\ and\ \bibinfo {author} {\bibfnamefont {D.~E.}\ \bibnamefont
  {Logan}},\ }\href@noop {} {\bibfield  {journal} {\bibinfo  {journal} {The
  European Physical Journal B-Condensed Matter and Complex Systems}\ }\textbf
  {\bibinfo {volume} {39}},\ \bibinfo {pages} {313} (\bibinfo {year}
  {2004})}\BibitemShut {NoStop}%
\bibitem [{\citenamefont {Vidhyadhiraja}\ and\ \citenamefont
  {Logan}(2005)}]{vidhyadhiraja2005optical}%
  \BibitemOpen
  \bibfield  {author} {\bibinfo {author} {\bibfnamefont {N.}~\bibnamefont
  {Vidhyadhiraja}}\ and\ \bibinfo {author} {\bibfnamefont {D.~E.}\ \bibnamefont
  {Logan}},\ }\href@noop {} {\bibfield  {journal} {\bibinfo  {journal} {Journal
  of Physics: Condensed Matter}\ }\textbf {\bibinfo {volume} {17}},\ \bibinfo
  {pages} {2959} (\bibinfo {year} {2005})}\BibitemShut {NoStop}%
\bibitem [{\citenamefont {Logan}\ and\ \citenamefont
  {Vidhyadhiraja}(2005)}]{logan2005dynamics}%
  \BibitemOpen
  \bibfield  {author} {\bibinfo {author} {\bibfnamefont {D.~E.}\ \bibnamefont
  {Logan}}\ and\ \bibinfo {author} {\bibfnamefont {N.}~\bibnamefont
  {Vidhyadhiraja}},\ }\href@noop {} {\bibfield  {journal} {\bibinfo  {journal}
  {Journal of Physics: Condensed Matter}\ }\textbf {\bibinfo {volume} {17}},\
  \bibinfo {pages} {2935} (\bibinfo {year} {2005})}\BibitemShut {NoStop}%
\bibitem [{\citenamefont {Logan}\ \emph {et~al.}(1998)\citenamefont {Logan},
  \citenamefont {Eastwood},\ and\ \citenamefont {Tusch}}]{logan1998local}%
  \BibitemOpen
  \bibfield  {author} {\bibinfo {author} {\bibfnamefont {D.~E.}\ \bibnamefont
  {Logan}}, \bibinfo {author} {\bibfnamefont {M.~P.}\ \bibnamefont {Eastwood}},
  \ and\ \bibinfo {author} {\bibfnamefont {M.~A.}\ \bibnamefont {Tusch}},\
  }\href@noop {} {\bibfield  {journal} {\bibinfo  {journal} {Journal of
  Physics: Condensed Matter}\ }\textbf {\bibinfo {volume} {10}},\ \bibinfo
  {pages} {2673} (\bibinfo {year} {1998})}\BibitemShut {NoStop}%
\bibitem [{\citenamefont {Sch{\"o}tt}\ \emph {et~al.}(2016)\citenamefont
  {Sch{\"o}tt}, \citenamefont {Locht}, \citenamefont {Lundin}, \citenamefont
  {Gr{\aa}n{\"a}s}, \citenamefont {Eriksson},\ and\ \citenamefont
  {Di~Marco}}]{schott2016analytic}%
  \BibitemOpen
  \bibfield  {author} {\bibinfo {author} {\bibfnamefont {J.}~\bibnamefont
  {Sch{\"o}tt}}, \bibinfo {author} {\bibfnamefont {I.~L.}\ \bibnamefont
  {Locht}}, \bibinfo {author} {\bibfnamefont {E.}~\bibnamefont {Lundin}},
  \bibinfo {author} {\bibfnamefont {O.}~\bibnamefont {Gr{\aa}n{\"a}s}},
  \bibinfo {author} {\bibfnamefont {O.}~\bibnamefont {Eriksson}}, \ and\
  \bibinfo {author} {\bibfnamefont {I.}~\bibnamefont {Di~Marco}},\ }\href@noop
  {} {\bibfield  {journal} {\bibinfo  {journal} {Physical Review B}\ }\textbf
  {\bibinfo {volume} {93}},\ \bibinfo {pages} {075104} (\bibinfo {year}
  {2016})}\BibitemShut {NoStop}%
\bibitem [{\citenamefont {Vidhyadhiraja}\ \emph {et~al.}(2009)\citenamefont
  {Vidhyadhiraja}, \citenamefont {Macridin}, \citenamefont {{\c{S}}en},
  \citenamefont {Jarrell},\ and\ \citenamefont
  {Ma}}]{vidhyadhiraja2009quantum}%
  \BibitemOpen
  \bibfield  {author} {\bibinfo {author} {\bibfnamefont {N.}~\bibnamefont
  {Vidhyadhiraja}}, \bibinfo {author} {\bibfnamefont {A.}~\bibnamefont
  {Macridin}}, \bibinfo {author} {\bibfnamefont {C.}~\bibnamefont {{\c{S}}en}},
  \bibinfo {author} {\bibfnamefont {M.}~\bibnamefont {Jarrell}}, \ and\
  \bibinfo {author} {\bibfnamefont {M.}~\bibnamefont {Ma}},\ }\href@noop {}
  {\bibfield  {journal} {\bibinfo  {journal} {Physical review letters}\
  }\textbf {\bibinfo {volume} {102}},\ \bibinfo {pages} {206407} (\bibinfo
  {year} {2009})}\BibitemShut {NoStop}%
\bibitem [{\citenamefont {Chen}\ \emph {et~al.}(2012)\citenamefont {Chen},
  \citenamefont {Meng}, \citenamefont {Pruschke}, \citenamefont {Moreno},\ and\
  \citenamefont {Jarrell}}]{chen2012lifshitz}%
  \BibitemOpen
  \bibfield  {author} {\bibinfo {author} {\bibfnamefont {K.-S.}\ \bibnamefont
  {Chen}}, \bibinfo {author} {\bibfnamefont {Z.~Y.}\ \bibnamefont {Meng}},
  \bibinfo {author} {\bibfnamefont {T.}~\bibnamefont {Pruschke}}, \bibinfo
  {author} {\bibfnamefont {J.}~\bibnamefont {Moreno}}, \ and\ \bibinfo {author}
  {\bibfnamefont {M.}~\bibnamefont {Jarrell}},\ }\href@noop {} {\bibfield
  {journal} {\bibinfo  {journal} {Physical Review B}\ }\textbf {\bibinfo
  {volume} {86}},\ \bibinfo {pages} {165136} (\bibinfo {year}
  {2012})}\BibitemShut {NoStop}%
\bibitem [{\citenamefont {Pourovskii}\ \emph {et~al.}(2013)\citenamefont
  {Pourovskii}, \citenamefont {Miyake}, \citenamefont {Simak}, \citenamefont
  {Ruban}, \citenamefont {Dubrovinsky},\ and\ \citenamefont
  {Abrikosov}}]{pourovskii2013electronic}%
  \BibitemOpen
  \bibfield  {author} {\bibinfo {author} {\bibfnamefont {L.}~\bibnamefont
  {Pourovskii}}, \bibinfo {author} {\bibfnamefont {T.}~\bibnamefont {Miyake}},
  \bibinfo {author} {\bibfnamefont {S.}~\bibnamefont {Simak}}, \bibinfo
  {author} {\bibfnamefont {A.~V.}\ \bibnamefont {Ruban}}, \bibinfo {author}
  {\bibfnamefont {L.}~\bibnamefont {Dubrovinsky}}, \ and\ \bibinfo {author}
  {\bibfnamefont {I.}~\bibnamefont {Abrikosov}},\ }\href@noop {} {\bibfield
  {journal} {\bibinfo  {journal} {Physical Review B}\ }\textbf {\bibinfo
  {volume} {87}},\ \bibinfo {pages} {115130} (\bibinfo {year}
  {2013})}\BibitemShut {NoStop}%
\bibitem [{\citenamefont {Keen}\ \emph {et~al.}(2020)\citenamefont {Keen},
  \citenamefont {Maier}, \citenamefont {Johnston},\ and\ \citenamefont
  {Lougovski}}]{keen2020quantum}%
  \BibitemOpen
  \bibfield  {author} {\bibinfo {author} {\bibfnamefont {T.}~\bibnamefont
  {Keen}}, \bibinfo {author} {\bibfnamefont {T.}~\bibnamefont {Maier}},
  \bibinfo {author} {\bibfnamefont {S.}~\bibnamefont {Johnston}}, \ and\
  \bibinfo {author} {\bibfnamefont {P.}~\bibnamefont {Lougovski}},\ }\href@noop
  {} {\bibfield  {journal} {\bibinfo  {journal} {Quantum Science and
  Technology}\ }\textbf {\bibinfo {volume} {5}},\ \bibinfo {pages} {035001}
  (\bibinfo {year} {2020})}\BibitemShut {NoStop}%
\bibitem [{\citenamefont {Maier}\ \emph {et~al.}(2005)\citenamefont {Maier},
  \citenamefont {Jarrell}, \citenamefont {Pruschke},\ and\ \citenamefont
  {Hettler}}]{maier2005quantum}%
  \BibitemOpen
  \bibfield  {author} {\bibinfo {author} {\bibfnamefont {T.}~\bibnamefont
  {Maier}}, \bibinfo {author} {\bibfnamefont {M.}~\bibnamefont {Jarrell}},
  \bibinfo {author} {\bibfnamefont {T.}~\bibnamefont {Pruschke}}, \ and\
  \bibinfo {author} {\bibfnamefont {M.~H.}\ \bibnamefont {Hettler}},\
  }\href@noop {} {\bibfield  {journal} {\bibinfo  {journal} {Reviews of Modern
  Physics}\ }\textbf {\bibinfo {volume} {77}},\ \bibinfo {pages} {1027}
  (\bibinfo {year} {2005})}\BibitemShut {NoStop}%
\bibitem [{err()}]{errorbar}%
  \BibitemOpen
  \href@noop {} {}\bibinfo {note} {The error bar represents the standard error
  rendered by regression analysis of the fitting.}\BibitemShut {Stop}%
\bibitem [{\citenamefont {Logan}\ and\ \citenamefont
  {Galpin}(2015)}]{logan2015mott}%
  \BibitemOpen
  \bibfield  {author} {\bibinfo {author} {\bibfnamefont {D.~E.}\ \bibnamefont
  {Logan}}\ and\ \bibinfo {author} {\bibfnamefont {M.~R.}\ \bibnamefont
  {Galpin}},\ }\href@noop {} {\bibfield  {journal} {\bibinfo  {journal}
  {Journal of Physics: Condensed Matter}\ }\textbf {\bibinfo {volume} {28}},\
  \bibinfo {pages} {025601} (\bibinfo {year} {2015})}\BibitemShut {NoStop}%
\bibitem [{\citenamefont {Eckstein}\ \emph {et~al.}(2007)\citenamefont
  {Eckstein}, \citenamefont {Kollar},\ and\ \citenamefont {Vollhardt}}]{isb1}%
  \BibitemOpen
  \bibfield  {author} {\bibinfo {author} {\bibfnamefont {M.}~\bibnamefont
  {Eckstein}}, \bibinfo {author} {\bibfnamefont {M.}~\bibnamefont {Kollar}}, \
  and\ \bibinfo {author} {\bibfnamefont {D.}~\bibnamefont {Vollhardt}},\
  }\href@noop {} {\bibfield  {journal} {\bibinfo  {journal} {Journal of Low
  Temperature Physics}\ }\textbf {\bibinfo {volume} {147}},\ \bibinfo {pages}
  {279} (\bibinfo {year} {2007})}\BibitemShut {NoStop}%
\bibitem [{\citenamefont {Glossop}\ \emph {et~al.}(2011)\citenamefont
  {Glossop}, \citenamefont {Kirchner}, \citenamefont {Pixley},\ and\
  \citenamefont {Si}}]{glossop2011critical}%
  \BibitemOpen
  \bibfield  {author} {\bibinfo {author} {\bibfnamefont {M.~T.}\ \bibnamefont
  {Glossop}}, \bibinfo {author} {\bibfnamefont {S.}~\bibnamefont {Kirchner}},
  \bibinfo {author} {\bibfnamefont {J.~H.}\ \bibnamefont {Pixley}}, \ and\
  \bibinfo {author} {\bibfnamefont {Q.}~\bibnamefont {Si}},\ }\href {\doibase
  10.1103/PhysRevLett.107.076404} {\bibfield  {journal} {\bibinfo  {journal}
  {Phys. Rev. Lett.}\ }\textbf {\bibinfo {volume} {107}},\ \bibinfo {pages}
  {076404} (\bibinfo {year} {2011})}\BibitemShut {NoStop}%
\bibitem [{\citenamefont {Pixley}\ \emph {et~al.}(2013)\citenamefont {Pixley},
  \citenamefont {Kirchner}, \citenamefont {Ingersent},\ and\ \citenamefont
  {Si}}]{pixley2013quantum}%
  \BibitemOpen
  \bibfield  {author} {\bibinfo {author} {\bibfnamefont {J.}~\bibnamefont
  {Pixley}}, \bibinfo {author} {\bibfnamefont {S.}~\bibnamefont {Kirchner}},
  \bibinfo {author} {\bibfnamefont {K.}~\bibnamefont {Ingersent}}, \ and\
  \bibinfo {author} {\bibfnamefont {Q.}~\bibnamefont {Si}},\ }\href@noop {}
  {\bibfield  {journal} {\bibinfo  {journal} {Physical Review B}\ }\textbf
  {\bibinfo {volume} {88}},\ \bibinfo {pages} {245111} (\bibinfo {year}
  {2013})}\BibitemShut {NoStop}%
\bibitem [{\citenamefont {Vojta}\ and\ \citenamefont
  {Kir\ifmmode~\acute{c}\else \'{c}\fi{}an}(2003)}]{vojta_kircan_2003}%
  \BibitemOpen
  \bibfield  {author} {\bibinfo {author} {\bibfnamefont {M.}~\bibnamefont
  {Vojta}}\ and\ \bibinfo {author} {\bibfnamefont {M.}~\bibnamefont
  {Kir\ifmmode~\acute{c}\else \'{c}\fi{}an}},\ }\href {\doibase
  10.1103/PhysRevLett.90.157203} {\bibfield  {journal} {\bibinfo  {journal}
  {Phys. Rev. Lett.}\ }\textbf {\bibinfo {volume} {90}},\ \bibinfo {pages}
  {157203} (\bibinfo {year} {2003})}\BibitemShut {NoStop}%
\bibitem [{\citenamefont {Logan}\ \emph {et~al.}(2016)\citenamefont {Logan},
  \citenamefont {Galpin},\ and\ \citenamefont {Mannouch}}]{logan2016mott}%
  \BibitemOpen
  \bibfield  {author} {\bibinfo {author} {\bibfnamefont {D.~E.}\ \bibnamefont
  {Logan}}, \bibinfo {author} {\bibfnamefont {M.~R.}\ \bibnamefont {Galpin}}, \
  and\ \bibinfo {author} {\bibfnamefont {J.}~\bibnamefont {Mannouch}},\
  }\href@noop {} {\bibfield  {journal} {\bibinfo  {journal} {Journal of
  Physics: Condensed Matter}\ }\textbf {\bibinfo {volume} {28}},\ \bibinfo
  {pages} {455601} (\bibinfo {year} {2016})}\BibitemShut {NoStop}%
\bibitem [{\citenamefont {Held}\ and\ \citenamefont
  {Bulla}(2000)}]{held2000mott}%
  \BibitemOpen
  \bibfield  {author} {\bibinfo {author} {\bibfnamefont {K.}~\bibnamefont
  {Held}}\ and\ \bibinfo {author} {\bibfnamefont {R.}~\bibnamefont {Bulla}},\
  }\href@noop {} {\bibfield  {journal} {\bibinfo  {journal} {The European
  Physical Journal B-Condensed Matter and Complex Systems}\ }\textbf {\bibinfo
  {volume} {17}},\ \bibinfo {pages} {7} (\bibinfo {year} {2000})}\BibitemShut
  {NoStop}%
\bibitem [{\citenamefont {Eisenlohr}\ \emph
  {et~al.}(2019{\natexlab{b}})\citenamefont {Eisenlohr}, \citenamefont {Lee},\
  and\ \citenamefont {Vojta}}]{eisenlohr2019mott}%
  \BibitemOpen
  \bibfield  {author} {\bibinfo {author} {\bibfnamefont {H.}~\bibnamefont
  {Eisenlohr}}, \bibinfo {author} {\bibfnamefont {S.-S.~B.}\ \bibnamefont
  {Lee}}, \ and\ \bibinfo {author} {\bibfnamefont {M.}~\bibnamefont {Vojta}},\
  }\href@noop {} {\bibfield  {journal} {\bibinfo  {journal} {Physical Review
  B}\ }\textbf {\bibinfo {volume} {100}},\ \bibinfo {pages} {155152} (\bibinfo
  {year} {2019}{\natexlab{b}})}\BibitemShut {NoStop}%
\bibitem [{\citenamefont {Glossop}\ \emph {et~al.}(2005)\citenamefont
  {Glossop}, \citenamefont {Jones},\ and\ \citenamefont
  {Logan}}]{glossop2005local}%
  \BibitemOpen
  \bibfield  {author} {\bibinfo {author} {\bibfnamefont {M.~T.}\ \bibnamefont
  {Glossop}}, \bibinfo {author} {\bibfnamefont {G.~E.}\ \bibnamefont {Jones}},
  \ and\ \bibinfo {author} {\bibfnamefont {D.~E.}\ \bibnamefont {Logan}},\
  }\href@noop {} {\bibfield  {journal} {\bibinfo  {journal} {The Journal of
  Physical Chemistry B}\ }\textbf {\bibinfo {volume} {109}},\ \bibinfo {pages}
  {6564} (\bibinfo {year} {2005})}\BibitemShut {NoStop}%
\bibitem [{\citenamefont {Withoff}\ and\ \citenamefont
  {Fradkin}(1990)}]{withoff1990phase}%
  \BibitemOpen
  \bibfield  {author} {\bibinfo {author} {\bibfnamefont {D.}~\bibnamefont
  {Withoff}}\ and\ \bibinfo {author} {\bibfnamefont {E.}~\bibnamefont
  {Fradkin}},\ }\href@noop {} {\bibfield  {journal} {\bibinfo  {journal}
  {Physical review letters}\ }\textbf {\bibinfo {volume} {64}},\ \bibinfo
  {pages} {1835} (\bibinfo {year} {1990})}\BibitemShut {NoStop}%
\bibitem [{\citenamefont {Fritz}\ and\ \citenamefont
  {Vojta}(2004)}]{fritz2004phase}%
  \BibitemOpen
  \bibfield  {author} {\bibinfo {author} {\bibfnamefont {L.}~\bibnamefont
  {Fritz}}\ and\ \bibinfo {author} {\bibfnamefont {M.}~\bibnamefont {Vojta}},\
  }\href@noop {} {\bibfield  {journal} {\bibinfo  {journal} {Physical Review
  B}\ }\textbf {\bibinfo {volume} {70}},\ \bibinfo {pages} {214427} (\bibinfo
  {year} {2004})}\BibitemShut {NoStop}%
\bibitem [{\citenamefont {Vojta}\ and\ \citenamefont
  {Bulla}(2001)}]{vojta2001kondo}%
  \BibitemOpen
  \bibfield  {author} {\bibinfo {author} {\bibfnamefont {M.}~\bibnamefont
  {Vojta}}\ and\ \bibinfo {author} {\bibfnamefont {R.}~\bibnamefont {Bulla}},\
  }\href@noop {} {\bibfield  {journal} {\bibinfo  {journal} {Physical Review
  B}\ }\textbf {\bibinfo {volume} {65}},\ \bibinfo {pages} {014511} (\bibinfo
  {year} {2001})}\BibitemShut {NoStop}%
\bibitem [{\citenamefont {Iizuka}\ \emph {et~al.}(2012)\citenamefont {Iizuka},
  \citenamefont {Kundu}, \citenamefont {Narayan},\ and\ \citenamefont
  {Trivedi}}]{iizuka2012holographic}%
  \BibitemOpen
  \bibfield  {author} {\bibinfo {author} {\bibfnamefont {N.}~\bibnamefont
  {Iizuka}}, \bibinfo {author} {\bibfnamefont {N.}~\bibnamefont {Kundu}},
  \bibinfo {author} {\bibfnamefont {P.}~\bibnamefont {Narayan}}, \ and\
  \bibinfo {author} {\bibfnamefont {S.~P.}\ \bibnamefont {Trivedi}},\
  }\href@noop {} {\bibfield  {journal} {\bibinfo  {journal} {Journal of High
  Energy Physics}\ }\textbf {\bibinfo {volume} {2012}},\ \bibinfo {pages} {1}
  (\bibinfo {year} {2012})}\BibitemShut {NoStop}%
\bibitem [{\citenamefont {Si}\ \emph {et~al.}(2001)\citenamefont {Si},
  \citenamefont {Rabello}, \citenamefont {Ingersent},\ and\ \citenamefont
  {Smith}}]{si2001locally}%
  \BibitemOpen
  \bibfield  {author} {\bibinfo {author} {\bibfnamefont {Q.}~\bibnamefont
  {Si}}, \bibinfo {author} {\bibfnamefont {S.}~\bibnamefont {Rabello}},
  \bibinfo {author} {\bibfnamefont {K.}~\bibnamefont {Ingersent}}, \ and\
  \bibinfo {author} {\bibfnamefont {J.~L.}\ \bibnamefont {Smith}},\ }\href@noop
  {} {\bibfield  {journal} {\bibinfo  {journal} {Nature}\ }\textbf {\bibinfo
  {volume} {413}},\ \bibinfo {pages} {804} (\bibinfo {year}
  {2001})}\BibitemShut {NoStop}%
\bibitem [{\citenamefont {Mukherjee}\ and\ \citenamefont
  {Lal}(2020{\natexlab{a}})}]{mukherjee2020scaling1}%
  \BibitemOpen
  \bibfield  {author} {\bibinfo {author} {\bibfnamefont {A.}~\bibnamefont
  {Mukherjee}}\ and\ \bibinfo {author} {\bibfnamefont {S.}~\bibnamefont
  {Lal}},\ }\href@noop {} {\bibfield  {journal} {\bibinfo  {journal} {New
  Journal of Physics}\ }\textbf {\bibinfo {volume} {22}},\ \bibinfo {pages}
  {063008} (\bibinfo {year} {2020}{\natexlab{a}})}\BibitemShut {NoStop}%
\bibitem [{\citenamefont {Mukherjee}\ and\ \citenamefont
  {Lal}(2020{\natexlab{b}})}]{mukherjee2020scaling2}%
  \BibitemOpen
  \bibfield  {author} {\bibinfo {author} {\bibfnamefont {A.}~\bibnamefont
  {Mukherjee}}\ and\ \bibinfo {author} {\bibfnamefont {S.}~\bibnamefont
  {Lal}},\ }\href@noop {} {\bibfield  {journal} {\bibinfo  {journal} {New
  Journal of Physics}\ }\textbf {\bibinfo {volume} {22}},\ \bibinfo {pages}
  {063007} (\bibinfo {year} {2020}{\natexlab{b}})}\BibitemShut {NoStop}%
\end{thebibliography}%

\end{document}